\documentclass{article}

\usepackage[figuresright]{rotating}
\usepackage{amsmath,amsfonts, bm,multirow,comment,soul,bm, float,algorithm}
\usepackage{algpseudocode}
\usepackage[margin=0.8in,footskip=0.25in]{geometry}
\usepackage{mathtools}

\def\bSig\mathbf{\Sigma}

\newcommand{\te}{\bm{\theta}}

\newcommand{\F}{\bm{F}'}
\newcommand{\W}{\bm{W}}
\newcommand{\V}{\bm{V}}
\newcommand{\G}{\bm{G}}

\title{A Bayesian Non-Stationary Heteroskedastic Time Series Model for Multivariate Critical Care Data}

\author{Zayd Omar$\dagger$, David A. Stephens$\dagger$, Alexandra M. Schmidt*, David L. Buckeridge*}

\begin{document}
	\maketitle
	
	\begin{abstract}
		We propose a multivariate GARCH model for non-stationary health time series by modifying the variance of the observations of the standard state space model. The proposed model provides an intuitive way of dealing with heteroskedastic data using the conditional nature of state space models. We follow the Bayesian paradigm to perform the inference procedure. In particular, we use Markov chain Monte Carlo methods to obtain samples from the resultant posterior distribution. Due to the natural temporal correlation structure induced on model parameters, we use the forward filtering backward sampling algorithm to efficiently obtain samples from the posterior distribution.  The proposed model also handles missing data in a fully Bayesian fashion.  We validate our model on synthetic data, and then use it to analyze a data set obtained from an intensive care unit in a Montreal hospital.  We further show that our proposed models offer better performance, in terms of WAIC, than standard state space models. The proposed model provides a new way to model multivariate heteroskedastic non-stationary time series data and the simplicity in applying the WAIC allows us to compare competing models.
	\end{abstract}
	

	\noindent$\dagger$ Department of Mathematics and Statistics, McGill University, Montreal, Canada.\\
	\noindent$*$ Department of Epidemiology, Biostatistics and Occupational Health, McGill University, Montreal, Canada.\\
	
	\section{Introduction}
	\label{s:intro}
	Advances in medical instrumentation and technology have given researchers the ability to collect large amounts of patient data. In this paper we study patient data arising from multi-channel monitoring in intensive care units (ICUs). There is increasing amount of research \cite{stein2013challenges} \cite{sow2010real} that recognizes the importance and difficulty in analyzing real time data collected in the ICU. Modeling such data accurately would help to ensure better patient care and outcome.  Time series analysis of the heart rate (HR) allows for a non-invasive way to study the heart rate variability (HRV), which describes the variability of the heart rate of an individual over a period of time.  It is well known that the HRV is a reflection of various physiological factors that help modulate the rhythm of the heart. Many papers \cite{saykrs1973analysis, malik1998heart, grogan2004reduced, acharya2006heart} discuss the significance of HRV in preventing different cardio-vascular pathologies. This type of data usually resembles long structured time series which may be correlated with other physiological processes. Such data, while opening up many new avenues of research, has complicated the nature of the analysis and there is a need to develop suitable multivariate models for these processes.

	The time dependent structure of the data naturally leads to the usage of multivariate time series analyses and state space modeling. Various models have been developed for modeling health time series data \cite{acharya2006heart, baselli1987heart, jung2015implications, montanoHR,baselli1986spectral}, but a major complication is that many series exhibit non-linear and non-stationary dynamics; there is an increasing amount of research into modeling \cite{ghassemi2015multivariate} some of these additional complexities. One way of dealing with non-stationary data is to use switching linear dynamical systems to analyze multivariate data and estimate based on approximately stationary intervals\cite{li2016model}.

	Predicting various outcomes, such as mortality, as well as future evolution of the time series themselves, is also of great interest.  Recently, machine learning researchers have tackled the prediction problem using EHR data and have provided a great deal of flexibility in terms of modeling the data. Deep learning, recurrent neural networks, deep neural networks, support vector machines, random forests, have all been applied for a variety of clustering, prediction and modeling tasks based on vital signs data\cite{alloghani2020prospects, luo2016predicting, che2018recurrent, shillan2019use, johnson2016machine}.  These methods, however, have several limitations and drawbacks, for example when dealing with missing data\cite{che2018recurrent, sun2020review}. While some machine learning models have had a degree of success in predicting outcomes in some settings, these models are typically hard to interpret and their robustness to variations in the data is unclear.  It is also not easy to understand how these models represent uncertainty in the data. More concerning is that there seems to be a large range in the capability of these models to reproduce their results \cite{johnson2017reproducibility}. Finally, it is not entirely clear that many of these models respect the chronology of events provided by a time series model, instead converting the time series data into some form of static patient data\cite{caballero2015dynamically}.

\begin{figure}[H]
		\centerline{\includegraphics[width = 1\textwidth]{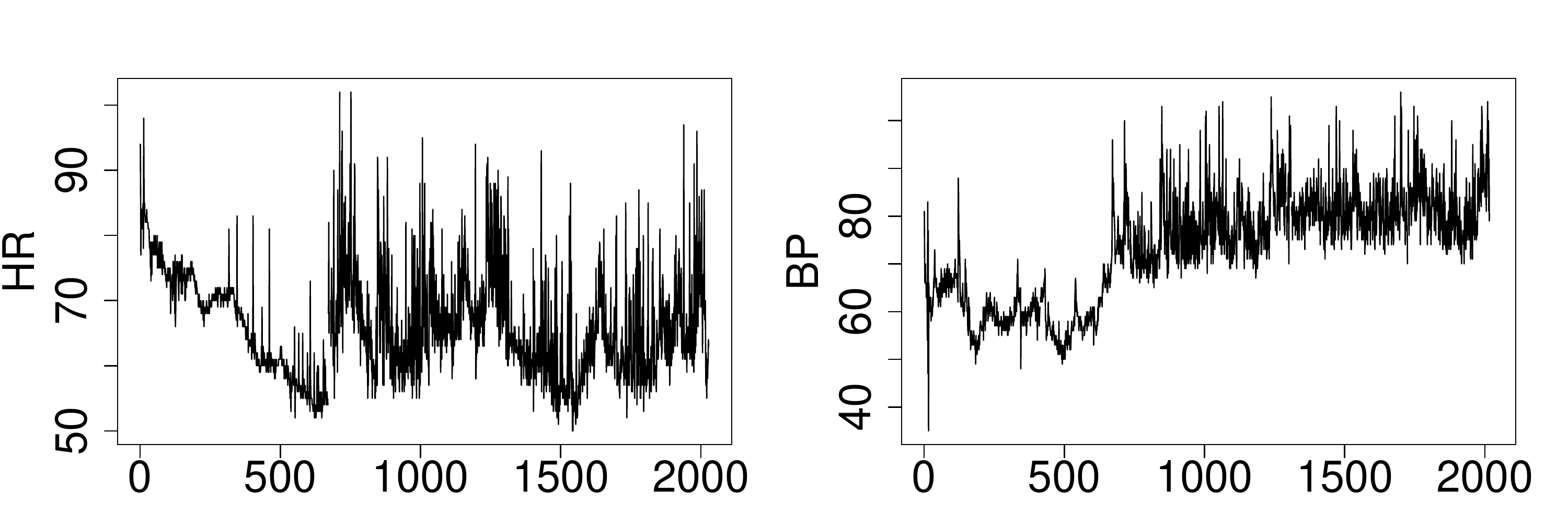}}
		\caption{Montreal ICU data for a single patient. Left: Heart rate series of a patient; Right: Blood pressure series of same patient. Time axis is number of seconds since start of monitoring\label{HrBpPlot}}
	\end{figure}

	Our proposed model and analysis are inspired by data obtained from an intensive care unit (ICU) at the McGill University Health Centre (MUHC), Montreal.  This motivated us to investigate non-stationary multivariate models for HR and BP. The data set records a large selection of physiological features collected from patients during their stay in the ICU and the readings of each patient are taken at 5 minute intervals over several days. The volume of measurements for each patient varies based on the duration of their stay at the hospital, and as a result for some patients there is a large volume of data while for others there is relatively little. Additionally, data are missing at random (due to various reasons, such as equipment failure).  Figure~\ref{HrBpPlot} shows two series, one for the HR and the second for BP, for a particular patient.
	
	The HR and BP series in Figure~\ref{HrBpPlot} are globally non-stationary and heteroskedastic. We are interested in building an interpretable and flexible model to fit the data. A joint model for this data would require modeling its mean-variance structure as well as the correlation structure. We study the HR and BP data under the framework of dynamic linear models (DLM) proposed by West and Harrison\cite{west2006bayesian}, which are a class of multivariate state-space models (SSMs) estimated under a Bayesian framework. SSMs allow for the modeling of non-stationary temporal data. We augment these dynamic models with conditionally heteroskedastic error terms, thus modeling the variance level dynamically together with the mean level of the time series. The main elaboration deploys the generalized autoregressive conditional heteroskedasticity (GARCH) model \cite{GARCH}.

	The proposed model leverages the conditional nature of both DLMs and GARCH type heteroskedasticity. This allows for simultaneous inference on the dynamic volatility of the process and accommodates the non-stationarity in the data, i.e. we can draw inference on time-evolution of the mean level of the HR and BP series and also get estimates on how the variance of each series is evolving. A natural consequence of our proposed model is that the inference procedure allows us to identify possible structural breaks in the data, places where there may have been an increased volatility (we discuss this in the data analysis part in Section 5). Our proposed model is also able to deal with missing data, which provides a significant advantage when using data sets like the one presented here, where missingness occurs at random. Finally, to further test the validity of our model, we use a 4-dimensional GARCH model to estimate data obtained from the MIMIC-III Waveform Database \cite{johnson2016db,johnson2016}. The MIMIC-III Waveform Database contains digitized ICU data for approximately 30,000 ICU patients. In particular we give an example of a joint modeling of HR, systolic BP, diastolic BP and respiration series.
	
	The proposed model is novel in the way that it uses conditional heteroskedasticity to explain the persistence of large variations observed in the patient data and focuses on the reconstruction and prediction of the time series values themselves. In the Bayesian setting, with large time series, prediction has its unique challenges \cite{aminikhanghahi2017survey}. There is a vast set of previous research that looks at the Bayesian prediction problem for both the offline and online settings \cite{adams2007bayesian, liu2020sequential, garnett2010sequential, ruggieri2016exact}. The proposed model is general enough to allow for the incorporation of such predictions, as one need only specify an appropriately tailored structure for the SSM -- see Section \ref{s:discuss} for an elaboration of the model designed to handle these cases.
	
	The patient-level analysis incorporated in this paper exclusively looks at a retrospective analysis of the patient level data instead of a predictive analysis so as to investigate the performance of the methodology in its ability to model the non-stationarity and conditional heteroskedasticity present in the data.  This investigation is also exclusively built to estimate the individual patient-level data over a population-based model. This, in the case of the ICU, appears more pertinent, as each patient admitted to the ICU is done so for their unique reasons, and the utility of a population-based analysis is not immediately apparent.  Nevertheless, our analysis can quite readily be extended to involve the building of a hierarchical population-based model, with an appropriate collection of hyper-prior specifications.  We discuss this briefly in Section \ref{s:discuss}.

	Overall, the proposed model is for non-linear and heteroskedastic time series data. It is easily interpretable and allows for predictions and population level estimation. It is able to deal with missing data, in an intuitive manner and does not impede analysis when the data are missing. Further, it is based on a more simple parameter space when compared to other machine learning models. Finally, in modeling the data with the proposed model, the results show the model is able to estimate the dynamic variance and the non-linear mean level of the model at each time period and investigate the big regime changes that occur during the patients' stay at the ICU.
	
	The paper is organized as follows. Section \ref{sec:model} gives a brief review of the theory of DLMs/SSMs and the GARCH processes, extends the standard Gaussian SSM to include GARCH errors and discusses model selection. Section \ref{s: Inference} develops an inferential procedure for the model. Section \ref{s:Simulation} describes a simulation study wherein we show that we are able to recover the values of the parameters. Section \ref{s:Application} applies the multivariate model to the MUHC and MIMIC-III data sets.  Section \ref{s:discuss} provides some discussion and presents future avenues of research.
	
	\section{Proposed Heteroskedastic Model}
	\label{sec:model}
	Our proposed heteroskedastic model combines the DLM and GARCH components to capture both non-stationarity in mean level and time-varying conditional variance. Suppose that $\{\bm{y}_t\}$ is a real-valued $(n\times 1)$ vector time series measured in discrete time. Let
	\begin{align}
		\begin{split}
			\text{Observation equation: } \qquad \bm{y}_t &= \bm{F}_t'\bm{\theta}_t+ \bm{z}_t,\\
			\text{State equation: } \qquad
			\bm{\theta}_t &= \bm{G}_{t}\bm{\theta}_{t-1}+\bm{w}_t, \label{DLM}
		\end{split}
	\end{align}
	with $\bm{w}_t \overset{i.i.d.}{\sim} N(\bm{0},\bm{W}_t)$. $\F_t$ is a known matrix $(n\times r)$ for all $t$. The residual error component, $\bm{z}_t$ $(n\times 1)$, is assumed to follow a multivariate GARCH process as defined by \cite{CCgarch} (details provided in Section 2.1).  The state equation describes a smooth evolution of the state parameter vector, $\te_t$ ($r  \times 1$), at time, $t$, as a function of the evolution matrix $\G_t$ ($r  \times r$), the previous values of the state vector plus some measurement error, $\bm{w}_t$ ($r  \times 1$). This measurement error is assumed to follow a multivariate normal distribution with zero mean and time-varying covariance matrix $\W_t$.  Let $\bm{V}_t$, an $(n\times n)$ matrix, be the variance of $\bm{z}_t$ (see  Section 2.1 for the formulation for the GARCH process). The residual errors of the observation and state equations, $\bm{z}_t$ and $\bm{w}_t$ respectively, are assumed independent across time. In this setup the unknown parameters are the variance parameters, that is, the components of $\bm{V}_t$ and $\W_t$. This formulation keeps the model very flexible and changing $\F_t$ and $\G_t$ provides many familiar models, including but not limited to seasonal models, linear growth models, linear regression models, and ARMA models.  A simple but surprisingly useful model is the `random-walk plus noise' (or `local level') model
	\begin{align}
		\begin{split}
			\text{Observation equation: } \qquad \bm{y}_t &= \bm{\theta}_t+ \bm{z}_t\\
			\text{State equation: } \qquad
			\bm{\theta}_t &= \bm{\theta}_{t-1}+\bm{w}_t\label{DLM-RWPN},
		\end{split}
	\end{align}
	with each equation in the same dimension, which can capture global non-stationarity.  In the state equation with $n=2$, we have
	\begin{align*}
		\theta_{1,t} & = \theta_{1,t-1} + w_{1j,t-1}\\
		\theta_{2,t} & = \theta_{2,t-1} + w_{2j,t-1}.
	\end{align*}
	More generally, the observation and state matrices may contain unknown parameters which may be inferred via the implied marginal likelihood for the observable quantities. When the observation and state level variances do not change with time, they are simply referred to as $\bm{V}$ and $\bm{W}$, respectively.
	
	The \textbf{dlm} package in \texttt{R} can fit the standard Bayesian DLM for many of the models indicated above, including models that exhibit mean-reversion, and by changing the form of the matrices $\bm{F}'$ and $\bm{G}$ at specific time points, can also fit models with structural breaks.  However, it does not include the option to fit with conditional heteroskedasticity, and as is observed in Figure \ref{HrBpPlot}, there is often visual evidence that time-varying conditional variability is present in the data.  In the next section, we describe the specific heteroskedastic model that we use.

	\subsection{Multivariate GARCH model}
	
	In \eqref{DLM}, $\bm{z}_t$ is assumed to be a Constant Conditional Correlation GARCH process (CCC-GARCH) which was proposed by \cite{CCgarch}, as given by the following formulation: we have that $\bm{z}_t = \bm{V}^{1/2}_t\bm{\epsilon_t}$, where $\bm{\epsilon_t}\sim N(\bm{0},\bm{I}_n)$ are i.i.d. and $\bm{V}_t^{1/2} = \textrm{chol}(\bm{V}_t)$, where
	\begin{equation}\label{GarchErr}
		\bm{V}_t  = \bm{D}_t\bm{R}\bm{D}_t',
	\end{equation}
	with $\bm{D_t} = \textrm{diag}(\sigma_{1,t},\dots,\sigma_{n,t})$ and, for $i=1,\dots,n$,
	\[
	\sigma_{i,t}^2 = \alpha_0^{(i)}+\alpha_1^{(i)}z_{i,t-1}^2+\dots+\alpha_p^{(i)}z_{i,t-p}^2 + \beta_1^{(i)}\sigma_{i,t-1}^2+\dots\beta_q^{(i)}\sigma_{i,t-q}^2
	\]
	with $\alpha_{0}^{(i)},\dots,\alpha_{p}^{(i)},\beta_{1}^{(i)},\dots,\beta_{q}^{(i)} \in \mathbb{R}_+^n$, and
	\begin{align}
		\bm{R} &= \begin{bmatrix}
			1 & \rho_{1,2} & \dots & \rho_{1,n}\\
			\rho_{2,1} & 1 & \dots & \rho_{2,n}\\
			\vdots & & \ddots & \vdots\\
			\rho_{n,1} & \rho_{n,2} & \dots & 1\label{RTmatrix} \\
		\end{bmatrix}.
	\end{align}
	Here $\textrm{chol}(\bm{V}_t)$ is the Cholesky decomposition of the matrix $\bm{V}_t$. In this model the correlations, $\rho_{i,j}$, in the symmetric matrix $\bm{R}$, remain constant over time. This correlation $\rho_{i,j}$ is the correlation between each series $z_{it}$ and $z_{jt}$. The GARCH parameters, $\alpha_{0}^{(i)},\dots,\alpha_{p}^{(i)},\beta_{1}^{(i)},\dots,\beta_{q}^{(i)}$ also must be placed under some joint constraint in order to maintain stationarity for the sequence $\{\bm{V}_t\}$. If all the GARCH parameters are zero, the model reverts back to a standard DLM. The correlation between each of the series given by $\rho_{i,j}$ for $i,=1,\dots,n$, is also a parameter of interest.

	The multivariate GARCH process $\bm{z}_t$ follows a martingale difference sequence and hence we have that, $E[\bm{z}_t] = 0, \; E[\bm{z}_t\bm{z}_s] = 0, \; t\neq s$. The simplest case for this model is when there is full independence of the time series at both the state level and the observation level and independence between the parameters. In this case each series can be fitted separately. However, when there exists correlation between the series,  fitting these time series individually will not be appropriate and a multivariate model should be fitted in order to appropriately account for the correlation present. In \eqref{RTmatrix}, the formulation that $\rho_{i,j}=0$ for all $i,j$ ($\bm{R}$ is then the $(n\times n)$ dimensional identity matrix) corresponds to the series being independent at the observation level. When $\rho_{i,j}$ is not zero and is unknown, the parameter space increases quadratically. For example if $\bm{z}_t$ is $(n\times 1)$ dimensional multivariate GARCH$(1,1)$ then the unknown parameters are $\alpha^{(i)}_0,\alpha^{(i)}_1,\beta^{(i)}_1$ $(i=1\dots n)$ and $\{\rho_{i,j}\}_{j,i=1,j> i}^n$, which gives the total number of unknown parameters as $3n+n(n-1)/2$. In the bivariate and trivariate cases, there are $7$ and $12$ unknown parameters respectively.
	
	\subsection{Mean-covariance structure and model properties}
	Let $D_t=\{\te_{0:t},\bm{y}_{1:t-1} \}$ be the set that contains the sequence of the state vector up to time $t$. Then $E[\bm{y}_t|D_t] = \F_t\te_t$ and $Var(\bm{y}_t|D_t) = \bm{V}_t$, where $\bm{V}_t$ is given in \eqref{GarchErr}. This formulation is similar to a standard GARCH process, where the variance at time $t$ depends conditionally on the square of the noise and the variance at time $t-1$. This allows for a dynamic variance at the observation level, thus extending the standard homoskedasticity assumption used in state-space models. As mentioned above, specifying different design matrices $\F$ and $\G$ allows for the building of many commonly used stationary and non-stationary state space models and these will be more flexible than the standard DLM due to the presence of a dynamic variance. This will later be seen in the bivariate model for HR and BP.
	
	\section{Bayesian inference procedure}\label{s: Inference}
	Let $\bm{y}_1,\dots,\bm{y}_{T}$ be a set of observed multivariate time series. Let $\bm{\psi}=\{\bm{\phi},\te_{0:T}\} $, where $\bm{\phi}$ is the set of unknown parameters ($\bm{\alpha}^{(1:n)}_{0:p},\bm{\beta}^{(1:n)}_{1:q},\bm{R},\bm{W}_{0:T}$), $\bm{\alpha}^{(i)}_{0:p}$ and $\bm{\beta}^{(1)}_{1:q}$ are the collection of all the GARCH parameters $\alpha^{(i)}_{0},\dots,\alpha^{(i)}_{p},\;\beta^{(i)}_1,\dots,\beta^{(i)}_{q}$, for each $i=1,\dots,n$.  Under the assumption that the residual errors arise from independent processes, and conditional on $\te_{0:T}$, the $\bm{y}_t$ are independent and the likelihood function is given by,
	\begin{align}
		\begin{split}
			\ell(\bm{y}_1,\dots,\bm{y}_T|\bm{\psi}) = \prod_{t = 1}^{T}f_{\bm{Y}}(\bm{y}_t|\bm{\psi})= f_{\bm{Y}}(\bm{y}_1|\bm{\psi}) \prod_{t = 2}^{T}f_{\bm{Y}}(\bm{y}_t|\bm{y}_{t-1},\bm{\psi}),
		\end{split}
	\end{align}
	where $f_{\bm{Y}}(\bm{y}_t|\bm{\psi})$ is a multivariate normal density with mean $\te_t$ and variance $\bm{V}_t$. Inference procedure is performed under the Bayesian paradigm, which naturally accounts for the uncertainty in the estimation of the parameters. The complete specification of the model under the Bayesian framework requires specifying an appropriate prior distribution $p(\bm{\phi})$ for $\bm{\phi}$.  Following Bayes' theorem the resultant posterior distribution for $\phi$ is proportional to,
	\begin{align}
		\begin{split}
			p(\bm{\phi}) f_{\bm{Y}}(\bm{y}_1|\bm{\phi}) \prod_{t = 2}^{T}f_{\bm{Y}}(\bm{y}_t|\bm{y}_{t-1},\bm{\phi}).
		\end{split}
	\end{align}
	We seek efficient computational approaches to compute the posterior for $\bm{\phi}$ as well as to infer the states in the latent model.
	
	\subsection{Kalman filter formulation}
	
	Despite the addition of the GARCH component, the proposed model can still be handled using techniques for the dynamic linear state-space model; see \cite{west2006bayesian} for further details.  We have by the Kalman recursions that $f_{\bm{Y}}(\bm{y}_t|\bm{y}_{t-1},\bm{\phi})$ is a multivariate Gaussian, with mean $\bm{f}_t$ and variance $\bm{Q}_t$, where
	\begin{align}
		\begin{split}
			\bm{f}_t = \F \bm{a}_t, \qquad \bm{Q_t} &= \bm{S}^2_t+\F\bm{H}_t\bm{F},\label{OneStepForeEqn}
		\end{split}
	\end{align}
	and we have
	\begin{align}\label{eq: filter_updates}
		\bm{a}_{t} & = \bm{G}_t\bm{m}_{t-1} \\[6pt]
		\bm{V}_{t} & =\bm{G}_t\bm{C}_{t-1}\bm{G}_t' +\bm{W}_t,
	\end{align}
	where the filtered values are, $\bm{m}_t=\bm{a}_t+\bm{K}_t\bm{e}_t$ and $\bm{C}_t=\bm{V}_t-\bm{K}_t\bm{Q}_t\bm{K_t}'$, with the Kalman gain and the forecast error being, $\bm{K}_t=\bm{V}_t\bm{F}_t\bm{Q}_t^{-1}$ and $\bm{e}_t = \bm{y}_t-\bm{f}_t$ respectively.
	
	The final element to complete the computation for \eqref{OneStepForeEqn} is the evaluation of the matrix $\bm{S}_t^2 = \bm{D}_t\bm{R}\bm{D}_t'$ where $\bm{D_t} = \textrm{diag}(S_{1,t},\dots,S_{n,t})$ and $\bm{R}$ is as defined in equation \eqref{RTmatrix} above.  Specifically, we have that
	\[
	S_{i,t}^2 = \alpha^{(i)}_0+ \sum_{j=1}^p \alpha^{(i)}_j \Big(y_{i,t-j}^2-2 y_{i,t-j}(\F\bm{m}_{t-j})_{(i)}+[\F(\bm{C}_{t-j}+\bm{m}_{t-j}\bm{m}_{t-j}')\bm{F}]_{(i,i)}\Big) + \sum_{j=1}^q \beta_j S^2_{i,t-j}.
	\]
	Where, $(i)$ represents the $i^{th}$ component of the vector and the $(i,i)$ represents the $i^{th}$ diagonal entry of the matrix. The marginal distributions for the one-step ahead forecasts are given by
	\[
	p(\bm{y}_{1:T}) = \int \prod_{t = 2}^{T}f_{\bm{Y}}(\bm{y}_t|\bm{y}_{t-1},\bm{\phi})p(\bm{\phi}) d\bm{\phi}.
	\]
	Since there is no closed form for the posterior distribution we use MCMC methods to sample from it.  Using the Forward-Filter-Backward-Sampling (FFBS) algorithm proposed by \cite{carter1994gibbs} and \cite{fruhwirth1994data} we can sample $\bm{\theta}=\te_{1:T}$ from the full conditional posterior distribution for the latent states.  Given the latent states and the prior distributions,  the conditional posterior for all other parameters is non standard and to obtain samples from their posterior full conditionals the Metropolis-Hastings algorithm is used.  For the GARCH parameters, the stationarity constraint makes efficient sampling complex.  Furthermore, in the constant correlation model, there is a positive definiteness constraint on the matrix which also requires attention.

	\subsection{Reparametrization of the correlation matrix, \textbf{\textit{R}}}\label{Inference}
	In the CCC-GARCH model, the correlation matrix, $\bm{R}$, must be positive definite and have diagonal elements equal to 1. A simple yet effective reparametrization of the matrix makes its estimation much simpler as will be seen subsequently in the simulation part of Section \ref{s:Simulation}. The parametrization comes from the upper Cholesky decomposition of $\bm{R}$, which allows us to write $\bm{R}=\bm{U}\bm{U}^\top$, where $\bm{U}$ is an $n \times n$ upper  triangular matrix with positive diagonal entries. The restriction that $\bm{R}$ has diagonal elements equal to one means that the row vectors of $\bm{U}$ are unit vectors. This gives us the following matrix form for $\bm{U}$:
	\[		
	\bm{U}=
	\begin{pmatrix}
		u_{11} & u_{12} & \dots & u_{1n}\\
		0 & u_{22} & \dots & u_{2n}\\
		\vdots & \vdots & \ddots & \vdots \\
		0 & 0 &  \dots & u_{nn}\\
	\end{pmatrix}.
	\]
	Notice that $u_{nn}=1$ by our formulation of the rows of $U$ being unit vectors and $u_{ii}>0$ for all $i=1,\dots,n$.  This reparametrization avoids having to estimate the components of $\bm{R}$ from \eqref{OneStepForeEqn} and instead estimates the elements of $\bm{U}$ since this will still uniquely determine $\bm{R}$. Because of this uniqueness, subsequently we will refer to these $u_{i,j}$ as the correlation components of the GARCH process. For more details on fast sampling of correlation matrices, see \cite{cordoba2018fast}. The only exception in using this parametrization is when $n=2$, the bivariate case.  Since in this case the only unknown in $\bm{R}$ is the single correlation coefficient, $\rho_{1,2}$, which has the restriction of being in the range $[-1,1]$.

	\subsection{Missing data}
	
	It is often the case that health time series data are subject to missingness in one or more of the components.  An advantage of the Bayesian method over non-Bayesian or non-probabilistic methods is that missing data can be imputed coherently as part of the inference.  MCMC-based inference is especially well suited to overcome missingness in the outcome component of the DLM in \eqref{DLM}; any missing $\bm{y}_t$ value, or a component of the observation vector, is sampled as part of the MCMC, usually via a Gibbs sampler step.  This step is particularly straightforward in a conditional Gaussian model as in \eqref{DLM}.
	
	Formally, having a missing value at time $t$ is the same as setting some components of $\F_t$ to zero. Suppose that $\bm{y}_t$ is a $p\times 1$ dimensional vector. There are two possibilities that we need to specify. The first one is when information is missing across the entire vector at time $t$, i.e. all the components of $\bm{y}_t$ are missing. In this case, the observation matrix $\F_t$ is zero. This also results in the Kalman gain matrix to be a matrix of $\bm{0}$s. The filtered updates are given as,
	\begin{align*}
		\begin{split}
			\bm{m}_t &= \bm{a}_t\\
			\bm{C}_t &= \bm{V}_t\\
		\end{split}
	\end{align*}
	
	The second possibility is that only some of the observations at time $t$ are missing. That is, for $\bm{y}_t$, only some components are missing while the others are present. In this case, we have that the components in the observation matrix $\F_t$ that correspond to the missing values, are set to $0$. This results in the a Kalman gain matrix, whose columns, corresponding to the missing values, contain zeros and the rest of the filtering process follows straight-forwardly as before.

	\subsection{Model selection}
	A key feature of estimating parameters is also deducing redundancies in the models. Model selection allows us to choose the best model given a choice of competing models. In our case the two competing models, a Gaussian-SSM and a GARCH-SSM. One of the main tools for model selection is the information criterion, in particular  Akaike's information criteria (AIC), Bayesian (or Schwarz) Information Criteria (BIC) use the maximum likelihood estimates and the log predictive distribution. Both AIC and BIC penalize for model complexity. We refer the reader to \cite{kitagawaIC} for more details.

	Following \cite{gelman2013bayesian} we use an information criteria proposed by \cite{waic}, known as the Watanabe-Akaike Information criteria also known as the Widely-Applicable Information Criteria (WAIC) to select the model. As mentioned in \cite{gelman2013bayesian}, the WAIC encompasses a Bayesian approach to the out-of-sample prediction performance and assesses the performance of the model using the samples drawn from the posterior distribution. The WAIC is computed as $\text{LPPD}-p_{\text{WAIC}}$, where $\text{LPPD}$ is the {log pointwise predictive density and $p_{\text{WAIC}}$ is a bias correction:
		\begin{align*}
			\text{LPPD} &= \sum_{i=1}^{n}\log  \int p(y_i|\bm{\psi})dp_{post}(\bm{\psi})=\sum_{i=1}^{n}\Big(\log[ E_{post} p(y_i|\bm{\psi} ) ] \Big) \\[6pt]
			p_{\text{WAIC}} &= 2\sum_{i=1}^{n}\Bigg\{\Big(\log[ E_{post} p(y_i|\bm{\psi} ) ] \Big)-\Big(E_{post} \log[ p(y_i|\bm{\psi} ) ] \Big)\Bigg\}.
		\end{align*}
		A higher value of the WAIC indicates a better fit for the model.  In our analysis, we compute the WAIC for the model where the latent state components are marginalized out.
		
		\section{Simulation study }\label{s:Simulation}
		In this section we carry out a simulation study for an illustrative 4-d GARCH-SSM. We simulate a GARCH-SSM data using a sample size of $T=1000$ and we estimate the parameters using both the standard state space model approach and our GARCH-SSM formulation.  The observation matrix, $\F$, and state evolution matrix, $\G$, are constant throughout time and fixed as $\F = \G = \bm{I}_4$, where $\bm{I}_4$ is the $4 \times 4$ dimensional identity matrix.  With the same parameter values we simulate 100 replicate data sets from a 4-dimensional GARCH(1,1)-SSM, thus allowing us to assess the frequentist behaviour in the parameter estimates.   The model parameters are the unknown components of the variance, that is the state variance and the components of the GARCH(1,1) error. We seek to estimate these parameters, the latent states and an estimate of the observation variance. The results show that our proposed model is broadly able to retrieve the unknown parameters, estimate the latent state generating the observations and is able to correctly identify a GARCH-model from a standard-SSM one. The full scope of the simulation studies are detailed in the Supplementary Materials provided online. All codes and results are also available and provided in the online Supplementary Materials section.
		
		\subsection*{Choice of priors}
		To estimate the model using the Bayesian framework we need to specify the priors. For the GARCH parameter $\alpha_0^{(i)}$ we choose a half-Cauchy prior; for $(\alpha_1^{(i)},\beta_1^{(i)})$ we choose a joint half-Cauchy prior restricted to the region such that $\alpha_1^{(i)}+\beta_1^{(i)}<1$; for the correlation parameters, $u_{ii}$, we choose a half-Cauchy prior, for the remainder of the correlation parameters, $u_{ij}$, we choose a $N(0,1)$ prior and for the state covariance matrix, $\W$, we choose an Inverse-Wishart prior, $IW(10,10)$, where, $i,j=1,2,3,4$.  The choice of priors reflects the positivity restriction that we have made for the GARCH parameters  and the diagonal components of the correlation components. The half-Cauchy prior has a thicker tail than other priors such as the half-normal and this allows for larger values for $\alpha_0$ to occur more frequently. This has the effect of allowing the baseline variance for the GARCH process to be high. We have also made a stationarity restriction for the GARCH parameters as given by the support $\alpha_1^{(i)}+\beta_1^{(i)}<1$ for $i=1,2,3,4$. For the correlation parameters in the matrix $\bm{U}$, inference needs to be carried out on the unnormalized version of $\bm{U}$ which then can normalized so that $\bm{R}=\bm{UU}^\top$ has unit diagonal entries. For the diagonal entries of the unnormalized version of $\bm{U}$ we choose a half-Cauchy prior, for the remainder of the correlation parameters, $u_{ij}$, we choose a $N(0,1)$ prior. Our choice of the Inverse-Wishart prior for the state variance $\W$, allows for a conjugate analysis. However, for all other parameters in the model we obtain samples from the posterior distributions using the Metropolis-Hastings algorithm.

		\subsection{Estimation, prediction and residual analysis}
		For these models, a key interest, other than the parameter estimates, are the estimation of the latent state vector and the dynamic variance to allow the study of heteroskedasticity.  We obtain posterior samples of the latent state vector using the FFBS algorithm. Being able to estimate the latent state allows us to study the mean level of the model.  Equation \eqref{OneStepForeEqn} gives, conditional on the posterior samples of the parameters, the posterior estimates of the one-step ahead forecast, $\bm{f}_t$, and its variance, $\bm{Q}_t$.   Similarly being able to estimate the evolution of the variance over time will allows us to study what led to moments of higher or lower variance, which is of interest to researchers performing a retrospective analysis of the data. For illustrations and more details regarding the simulation study we refer the reader to the Supplementary Materials. We are able to accurately recover the latent state, the dynamic variance and the unknown parameters. We also do a residual analysis in the Supplementary Materials section, showing that the heteroskedasticity adjusted residuals return approximately standard-normal error terms as expected under the specifications given by (\ref{GarchErr}).

		\subsection{Model selection}
		A key objective of the simulation study is to assess whether, at the selected sequence length, our estimated models are correctly selected using the WAIC when the heteroskedasticity is in fact present in the data generating mechanism. In each simulation we calculated the WAIC using from both the GARCH-SSM estimation and the standard-SSM estimation. Since our models were originally generated using a multivariate GARCH(1,1) we would ideally want that our WAIC is always choosing the GARCH model over the standard-SSM. From our simulations we get that the WAIC for the GARCH model is always higher than the WAIC of the standard model, that is we are always accurately choosing the GARCH model over the standard state space model. More details regarding the calculation of the WAIC are provided in Supplementary Materials.
		
		\section{Application to real ICU Data}\label{s:Application}
		In this section we will apply our bivariate GARCH-SSM to multivariate ICU data.  This data set is a selected patient observed over $T=2029$ data points at a frequency of one measurement every five minutes.  This data set has missing values in both BP and HR series and also short segments where there are no observations in either series.
		
		\subsection{Montreal ICU data}\label{sec:Montreal}
		
		As seen from Figure \ref{HrBpPlot}, both these series show non-stationarity properties in mean and a time-varying variance.  There are also apparent outliers in both series, probably due to some physiological change occurring; this happens roughly at the same time for both series. Given the sudden change in volatility and what seems to be pockets of volatility clustered together, we believe that the proposed GARCH model would be an appropriate model; our GARCH-SSM is based on a bivariate GARCH($1,1$) structure.
		
		We compare the fit from our proposed model to the fit from a standard DLM, which we use as a benchmark. First we fit a standard DLM, specifically, the random walk plus noise model from equation \eqref{DLM-RWPN} with a constant conditional variance structure, in two dimensions.  We assume that the observation and state matrices are constant through time and we choose $\F$ and $\G$ both as the 2-dimensional identity matrix.  In this model our unknown parameters are the observation variance, $\V$, and the state variance, $\W$. For both $\V$ and $\W$ we specify a Inverse-Wishart prior $IW(10,10)$, which are conjugate priors and makes sampling from the posterior easier. Posterior samples were obtained by running the MCMC using four chains, each starting from different values. Each chain had a burn-in sample of 20,000 and subsequently every $50^{th}$ sampled value was stored. This resulted in a total sample size of 8,000 from the posterior distribution. We checked for convergence of our chains using the ergodic means.

		Next, we fit a bivariate model with GARCH errors. The main difference from the previous model is that the observation error, $\bm{z}_t$, follows a bivariate CCC-GARCH(1,1) process, more specifically, we assume
		\begin{align*}
			\sigma_{1,t}^2 & = \alpha_0^{(1)}+\alpha_1^{(1)}y_{1,t-1}^2 +\beta_1^{(1)}\sigma_{1,t-1}^2 \\[6pt]
			\sigma_{2,t}^2 & = \alpha_0^{(2)}+\alpha_1^{(2)}y_{2,t-1}^2 +\beta_1^{(2)}\sigma_{2,t-1}^2,
		\end{align*}
		where, $\alpha_0^{(i)},\alpha_1^{(i)},\beta_1^{(i)}$ are the components of the observation variance parameters, where $i=1,2$ (which correspond to HR and BP respectively). These need to be estimated as does, $\W$, the state variance parameter. Define $\rho_s$ as the correlation of the state errors, that is it is obtained by dividing the off-diagonal element of $\W$ (the covariance between the state errors) by the standard deviations of the state errors. For the GARCH-parameters we choose a half-Cauchy prior for $\alpha_0^{(i)}$. For $(\alpha_1^{(i)},\beta_1^{(i)})$ we choose a joint half-Cauchy prior restricted to the region such that $\alpha_1^{(i)}+\beta_1^{(i)}<1$. Define $\rho_{obs}$ as the correlation between the observation errors. For $\rho_{obs}$, we choose a Uniform(-1,1) prior and for the state covariance matrix, $\W$, we choose an $IW(10,10)$. The Inverse-Wishart prior for $\W$ in this case is also a conjugate prior. However, for all other parameters, as the posterior full conditionals are unknown we use steps of the Metropolis-Hastings algorithm, with random walk proposals. Posterior samples were obtained from starting 4 different chains from random initial values. We discarded the first 20,000 iterations as burn-in and we stored every $50^{th}$ samples and collected a total of 8,000 samples from the posterior distribution for each parameter. We again checked for the convergence of our chains using the ergodic mean.

		The WAIC values were -24676.52 for the standard state space model and -23958.42 for the GARCH-SSM. Since the GARCH-SSM has a higher WAIC than the standard state space model, thus we have that the GARCH-SSM is the best among the fitted models. Table \ref{SSMestHRBP}, shows the posterior parameter estimates from a standard bivariate DLM. We see that for $\rho_{obs}$, the $95\%$ posterior credible interval contains $0$. However, there seems to be a positive correlation between the state vectors as given by, $\rho_s$.

		%\begin{table}[h]	
		%	\begin{center}
			%		\begin{tabular}{cc}
				%			% \rowcolor{lightgray}\multicolumn{2}{c}{\underline{WAIC}}\\
				%			\rowcolor{lightgray} Std-SSM & GARCH-SSM \\
				%			-24676.52 & -23958.42\\
				%		\end{tabular}
			%		\vspace{0.2cm}\caption{WAIC calculated from the Standard-Gaussian SSM and the GARCH-SSM}\label{SSMvsGarchWaic}
			%	\end{center}
		%\end{table}	

		\begin{table}[H]
			\centering
			\def\~{\hphantom{0}}
			\begin{tabular*}{\columnwidth}{@{\extracolsep{\fill}}c@{\extracolsep{\fill}}l@{\extracolsep{\fill}}c@{\extracolsep{\fill}}c}
				% \hline	
				% \rowcolor{lightgray}\multicolumn{4}{c}{\underline{Standard SSM Model}}\\
				\multicolumn{4}{c}{\underline{Observation Level Parameter Estimates}} \\
				& $\bm{V}_{HR}$ & $\bm{V}_{BP}$& $\bm{\rho}_{obs}$\\ 
				\textbf{Post. est.} & 25.168  &  19.813  &  0.05  \\
				\text{s.d.} & (0.873)  &  (0.534)  &  (0.358)\\
				\hline
				\multicolumn{4}{c}{\underline{State Level Parameter Estimates}} \\
				& $\bm{W}_{HR}$ & $\bm{W}_{BP}$& $\bm{\rho}_s$ \\ 
				\textbf{Post. est.} &  1.287  &  0.93  &  0.15\\
				\text{s.d.} & (0.038)  &  (0.02)  &  (0.005)\\
				\hline
			\end{tabular*}\vskip6pt
			\vspace{0.2cm}\caption{Montreal ICU data: Parameter estimates of the bivariate Gaussian-SSM. Here $\rho_{obs}$ is the estimate of the correlation between the observation errors of the HRT series and the BP series and $\rho_s$ is the estimate of the correlation between the state errors of both series. The values displayed are the posterior means of the parameters and the values in the parenthesis are the posterior standard deviations.}\label{SSMestHRBP}
		\end{table}

		Table \ref{GarchEstHRBP} gives the parameter estimates of the bivariate GARCH-SSM. We have that for both the HR and BP series, the $\beta_1$ parameter estimate is quite high (the posterior mean is $0.691$ for the HR series and $0.747$ for the BP series). This shows that there is quite a high persistence in volatility. At the same time in terms of the correlation, at the observation level there seems to be a very small correlation with the posterior estimate being $\rho_{obs}=0.098$ with the $95\%$ posterior credible interval being $(0.05,0.14)$. The state level posterior estimation of the correlation is $\rho_s=0.069$ and the $95\%$ posterior credible interval is $(-0.03,0.19)$. Figure \ref{PostDynVar} (top row) shows the posterior estimates of the state for each of the HR and BP series along with the 95\% credible interval given by the dashed lines. The posterior estimates of the latent state have been superimposed over the true observations to show how the estimated level of the observations is changing over time.

		\begin{table}[H]
			\centering
			\def\~{\hphantom{0}}
			\label{t:tablefour}
			\begin{tabular*}{\columnwidth}{@{\extracolsep{\fill}}c@{\extracolsep{\fill}}c@{\extracolsep{\fill}}c@{\extracolsep{\fill}}c@{}c}
				% \hline
				% \rowcolor{lightgray}\multicolumn{5}{c}{\underline{GARCH(1,1)-SSM Model}} \\
				\multicolumn{5}{c}{\underline{Estimates for HR series}} \\
				& $\bm{\alpha}^{HR}_{0}$ & $\bm{\alpha}^{HR}_{1}$& $\bm{\beta}^{HR}_{1}$& $\bm{W}^{HR}$\\ 
				\textbf{Post. est.} & 1.109 & 0.305 & 0.691 & 0.623 \\
				\text{s.d.}& (0.197)& (0.028) & (0.028) & (0.091) \\
				CI. 95\%& (0.81,1.57 ) & (0.26, 0.36) & (0.63, 0.74) & (0.47, 0.82)  \\[12pt]
				\hline
				\multicolumn{5}{c}{\underline{Estimates for BP series}} \\[6pt]
				& $\bm{\alpha}^{BP}_{0}$ & $\bm{\alpha}^{BP}_{1}$& $\bm{\beta}^{BP}_{1}$& $\bm{W}^{BP}$ \\	
				\textbf{Post. est.} & 0.777 & 0.248 & 0.747 & 0.415 \\
				\text{s.d.}& (0.173) & (0.029) & (0.029) & (0.063)\\
				CI. 95\%& (0.52 ,1.19 ) & (0.20, 0.31) & (0.69, 0.80) & (0.31, 0.56) \\[6pt]
				\hline
				\multicolumn{5}{c}{\underline{Estimates of Correlation Parameters}} \\
				& $\bm{\rho}$ & $\bm{\rho}_s$ & & \\
				\textbf{Post. est.} & 0.098 & 0.069 && \\
				\text{s.d.}&(0.024) & (0.057) & & \\
				CI 95\%  & $(0.05,  0.14)$ & $(-0.03, 0.19)$ & &\\
				\hline
			\end{tabular*}\vskip6pt
			\caption{Montreal ICU data:  Bivariate GARCH(1,1)-SSM parameter estimates. $\rho$ is the estimate of the correlation between the observation errors of the HRT series and the BP series and $\rho_s$ is the estimate of the correlation between the state errors of both series. The estimates are the posterior mean values and in the parenthesis is below is their posterior standard deviations, and below that the $95\%$ credible interval.}\label{GarchEstHRBP}
		\end{table}
		
		Figure \ref{PostDynVar} (bottom row) provides the posterior estimate of the variance of the two different series over time. We see that in both the HR and BP series the notion of either series having a constant variance is likely inaccurate.  We see that, especially for the HR series, there are moments with large jumps in the conditional variance, followed by sharp dips.  For both the HR and BP series allowing a dynamic variance seems appropriate given the uneven fluctuations in the data. As a consequence of the posterior estimation of the variance, we  can thus also gauge and see if there were any large structural changes in the model. We use this to see that for both the HR and BP series, after about the $700^{th}$ observation there is a change in the mean level of the variance (this may be caused by a physiological change which then manifests itself in the HR and BP data). The persistence of volatility in both the series can be seen from the panels in Figure \ref{PostDynVar}.

		\begin{figure}[H]
			\begin{center}
				\includegraphics[width=1\textwidth,height=0.25\textwidth]{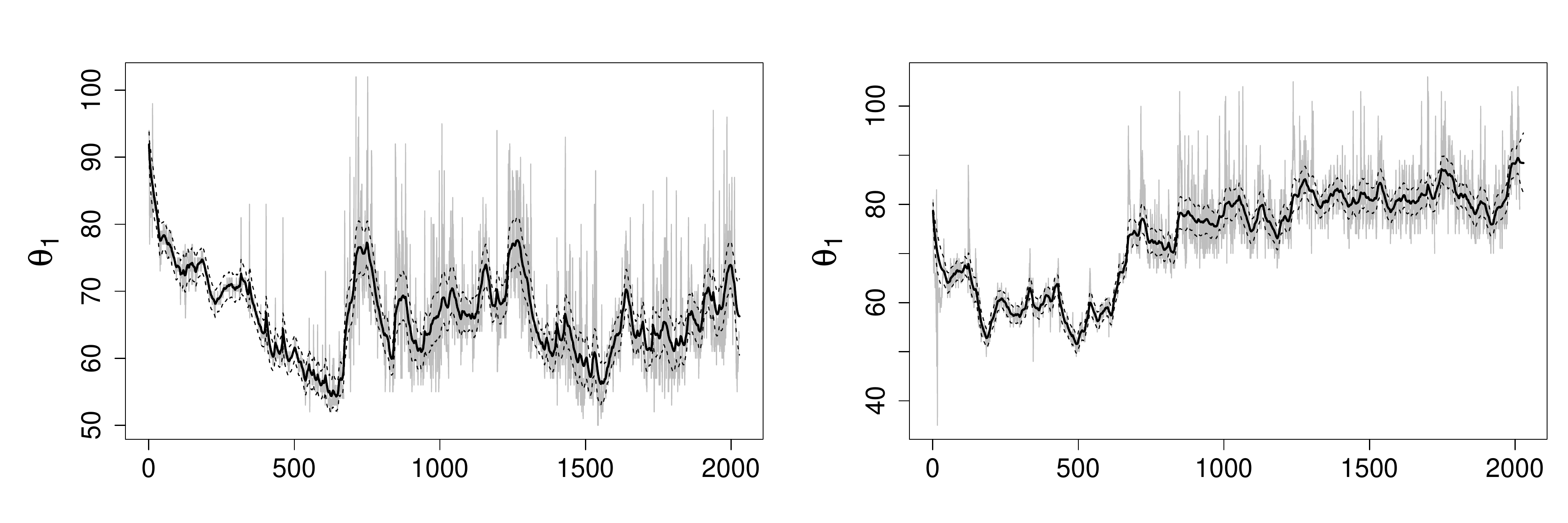}
				\includegraphics[width=1\textwidth,height=0.25\textwidth]{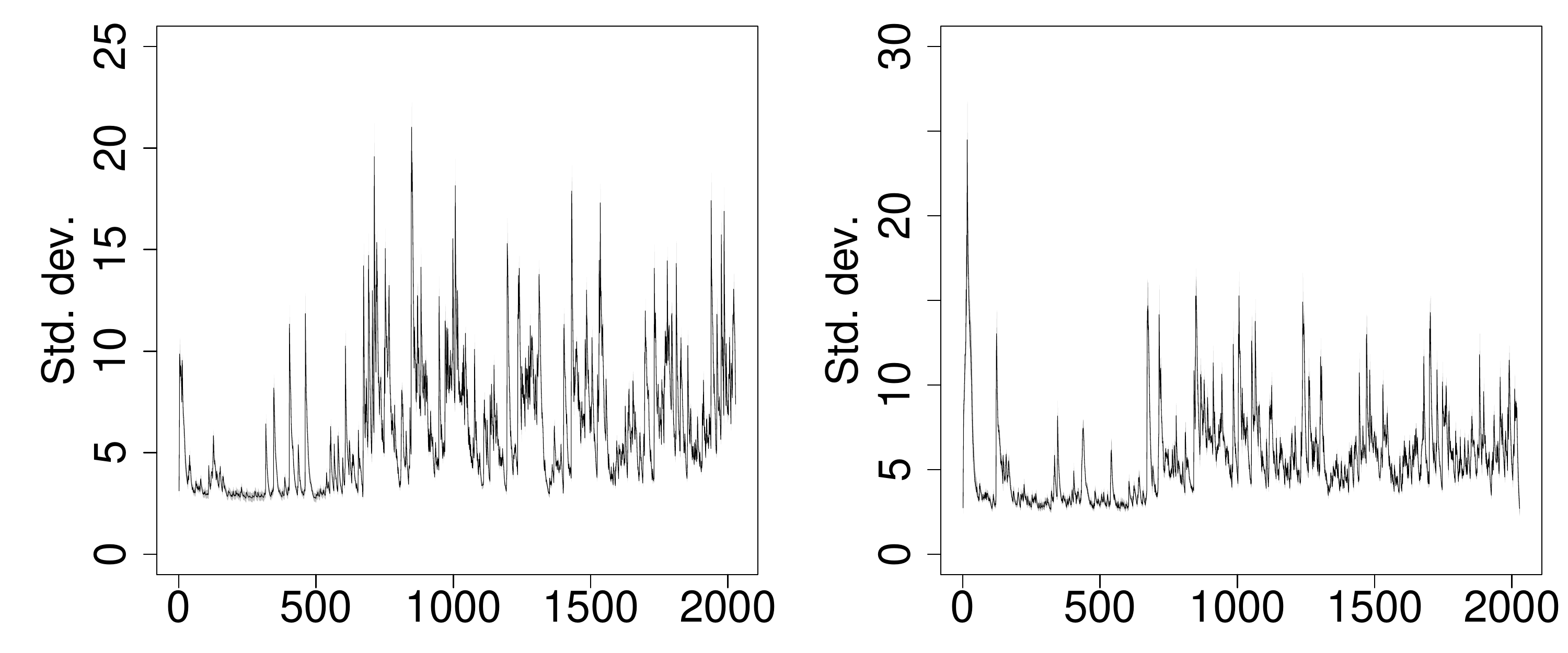}
			\end{center}
			\vspace{0.5cm}\caption{Montreal ICU data.  Top row: posterior estimates of the mean level of each series (HR left, BP right), with the credible intervals given by the dashed lines and the observed data in gray.  Bottom row: posterior mean estimates of the dynamic standard deviation (HR left, BP right).}\label{PostDynVar}
		\end{figure}
		
		In the analysis presented, the fit of the random-walk plus noise model was investigated, and the fit appears good.  However, it is also plausible that a model with a latent stochastic trend could be useful in explaining variation in critical care data.  In this model the componentwise state is augmented from $\theta_{j,t} = \mu_{j,t}$ (the local stochastic mean) to $\theta_{j,t} = (\mu_{j,t},\lambda_{j,t})^\prime$ (the local stochastic mean and trend), where
		\begin{align*}
			\theta_{j,t} & = \theta_{j,t-1} + \lambda_{j,t-1} + w_{j,t-1}\\
			\lambda_{j,t} & = \lambda_{j,t-1} + u_{2j,t-1},
		\end{align*}
		for $j=1,2$. This which can provide an additional smoothness to the reconstruction.  Further extensions would allow for structural breaks in the model and a time varying correlation between multiple series (recall that the proposed model keeps the correlation constant). However, fitting this results in an inferior WAIC compared to the random-walk noise model.

		\subsection{MIMIC ICU Data}\label{sec:MIMIC}
		We fit our 4-d GARCH model on patient data obtained from the MIMIC-III Waveform Database \cite{johnson2016db,johnson2016}.  We fit a joint model for HR, systolic BP, diastolic BP and respiration data for a single patient. Each series is complete, with no missingness, and has $T=647$ data points.  The model implementation was identical to that from section \ref{sec:Montreal}, namely the random-walk plus noise model, but now in four dimensions.  The MCMC implementation details were also identical. The parameter estimates of the model are presented in Table \ref{GarchEstMimic}.

		\begin{figure}[H]
			\begin{center}
				\includegraphics[width = 1.0\textwidth,height=0.3\textwidth]{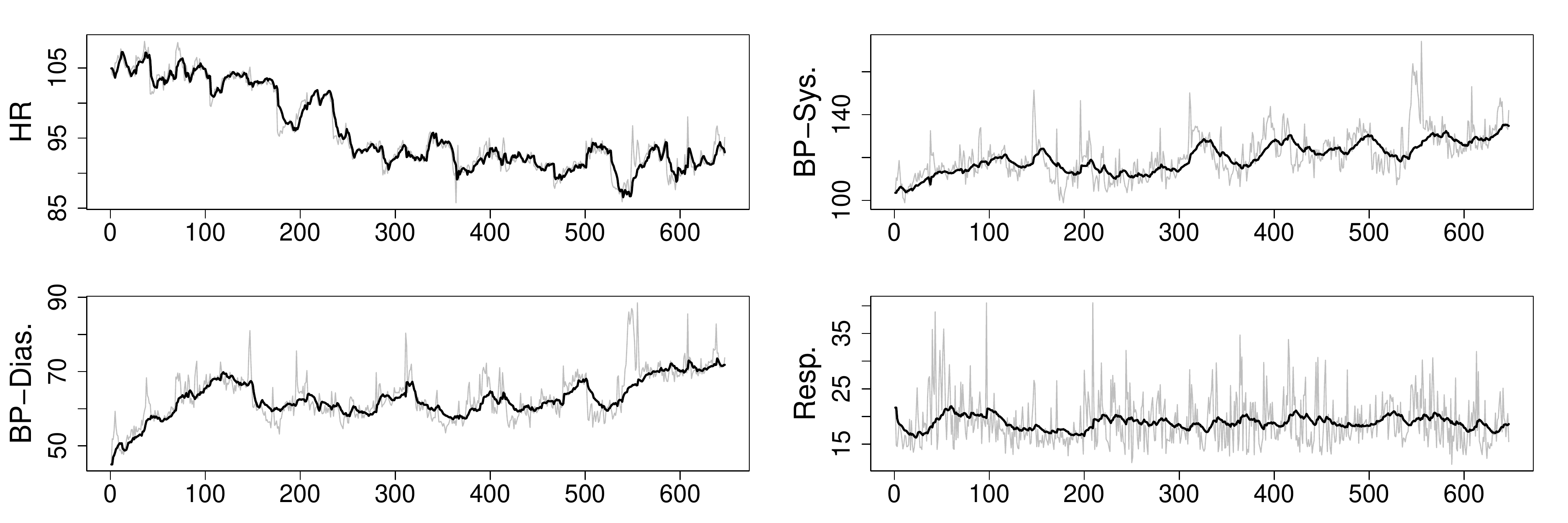}
			\end{center}
			\caption{MIMIC data: In black, the posterior estimates of the one step ahead forecasts of each series. In gray are the observed data.}\label{OneStepForeMimic}
			
			\begin{center}
				\includegraphics[width = 1.0\textwidth,height=0.3\textwidth]{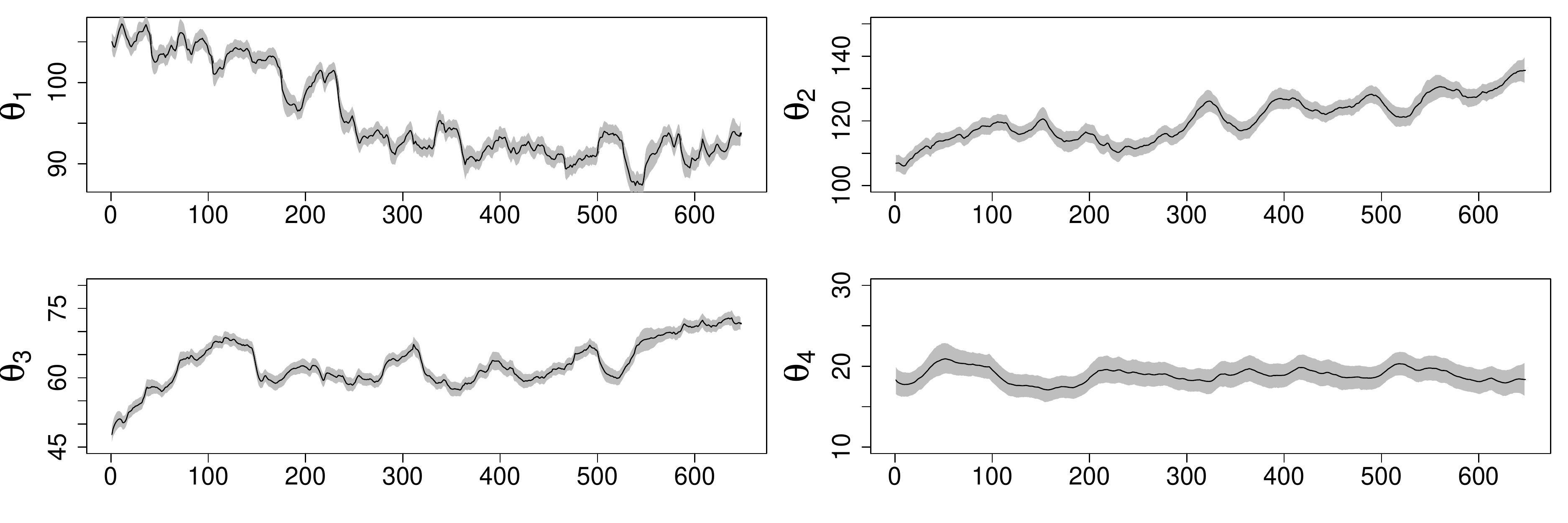}
			\end{center}
			\caption{MIMIC data: The thick dark line are the posterior median estimates of the states. The dashed lines are the true value of the states. The gray band is the 95\% posterior credible interval.}\label{PostStateMimic}
			\begin{center}
				\includegraphics[width = 1.0\textwidth,height=0.35\textwidth]{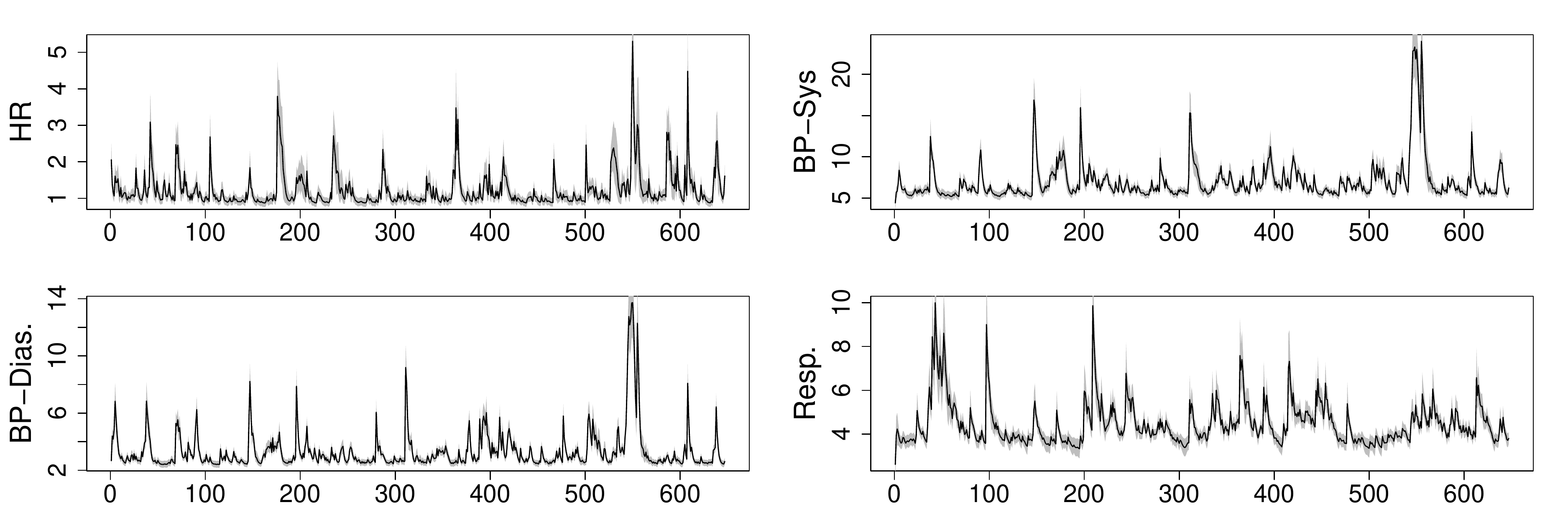}
			\end{center}
			\vspace{0.5cm}\caption{MIMIC data: The dynamic standard deviation in black. The shaded gray area is the 95\% posterior credible interval of the estimated standard deviation.}\label{DynVarMimic}
		\end{figure}

		\begin{table}[H]
			\centering
			\begin{tabular}{cccc}
				\multicolumn{4}{c}{\underline{\textbf{Parameter Estimates}}}\\
				& $\bm{\alpha}_{0}$ & $\bm{\alpha}_{1}$ & $\bm{\beta}_{1}$ \\
				\hline
				\multirow{1}{*}{\textbf{HR}} & $0.433$ & $0.360$ & $0.246$ \\
				\text{s.d.}  & (0.109) & (0.092) & (0.111) \\
				CI 95\% &  (0.25, 0.67) & (0.22, 0.58 ) & (0.054, 0.48 ) \\
				\hline
				\multirow{1}{*}{\textbf{BP-Sys}.} & $11.004$ & $0.196$ & $0.549$ \\
				\text{s.d.}  & (2.538) & 0.038 & 0.074 \\
				CI 95\% &  (7.06, 17.14 ) & (0.13, 0.28) & (0.39, 0.683) \\
				\hline
				\multirow{1}{*}{\textbf{BP-Dias}.} & $3.085$ & $0.265$ & $0.418$ \\
				\text{s.d.} & (0.513) & (0.047) & (0.061) \\
				CI 95\% &  (2.29, 4.30) & (0.19, 0.37 ) & (0.28, 0.53) \\
				\hline
				\multirow{1}{*}{\textbf{Respiration}} & $2.950$ & $0.153$ & $0.698$ \\
				\text{s.d.} & (1.067) & (0.050) & (0.078) \\
				CI 95\% &  (1.62, 5.80) & (0.080, 0.27) & (0.50, 0.82) \\
				\hline
			\end{tabular}
			\caption{MIMIC data: 4-d GARCH(1,1)-SSM parameter posterior median estimates  (Std. Error)}\label{GarchEstMimic}
		\end{table}
		
		The posterior estimate of the state variance matrix $\bm{W}$ and the GARCH correlation matrix $\bm{R}$ are respectively,
		\begin{align*}
			\begin{split}
				\begin{pmatrix*}[r]
					0.425 & -0.141 & -0.028  &0.004 \\
					- & 0.416  & 0.039 &-0.003\\
					- & - & 0.313 & -0.004\\
					- & - & - & 0.070
				\end{pmatrix*} \qquad
				\begin{pmatrix*}[r]
					1.00 & 0.44 & 0.48 & 0.25\\
					- & 1.00 & 0.92 & 0.31\\
					- & - & 1.00 & 0.33\\
					- & - & - & 1.00
				\end{pmatrix*}.
			\end{split}
		\end{align*}
		
		For this data set, the WAIC value of the GARCH model is -12404.2 while the WAIC using a DLM is -12440.92.  Thus in this case the GARCH model is preferred, and there is strong evidence for correlation in the GARCH errors, with estimated correlation as large as 0.92.  From Figure \ref{OneStepForeMimic} one can see that there are instances when the observed data seem to have some large fluctuations in variance. However, this data set is much more straightforward, and lacks the persistence of volatility present in the data from the Montreal ICU. Figure \ref{PostStateMimic} shows the posterior estimate of the state variable along with the $95\%$ credible intervals. This can further be seen from the estimates of the variance in this model given in Figure \ref{DynVarMimic} which shows that the variance is fairly stable with some predictable fluctuations over time. This also explains the values of the estimates of the GARCH parameters which are much far away from the non-stationary region for this model. Overall the parameter estimates from the model seem reasonable.
		
		\section{Discussion}\label{s:discuss}
		We proposed a new model to deal with non-stationary and heteroskedastic types of health time series. To the best of our knowledge this is the first type of model combining heteroskedasticity and non-stationarity for Bayesian state-space models in this manner and applying it to the health time series data. This is done by extending the classical DLM with a multivariate GARCH process. The structure of the mean-level model chosen was a mutlivariate random-walk plus noise model. A draw back of this structure is that it precludes any type of forward prediction estimates. However, with another type of mean-level chosen, the ability to forecast ahead is restored. The Bayesian algorithms developed are able to sample efficiently from the unknown posterior distribution using MCMC and they are able to satisfactorily recover the parameters of the model in simulation, the latent state and estimate the dynamic variance. We fitted the proposed model to the ICU dataset and showed an improvement in the prediction accuracy (as measured by WAIC) obtained using our proposed model over the standard multivariate SSMs currently available.
		
		In this paper, we have focussed on a state-space representation where the latent process is continuous valued, and represents the `noise-free' version of the patient-specific process for the observable quantities.  However, it is reasonably standard in the Bayesian literature to also allow for discrete states, in a hidden Markov model, that might be used to represent categorized health-states.  In the simplest case that might be useful for an ICU setting, we might propose a binary switching HMM where the states represent `stable' and `unstable' status of the patient, on top of which further latent and observable structures may be built.  The computation of the Bayesian posterior for such a model is more involved, but still tractable.  See the comprehensive coverage in Chapter 13 of \cite{Fruhwirth-Schnatter2006}, which provides several options for Markov switching formulations.
		
		We have studied only individual-level patient data as this is the most urgent issue in ICU care, but the Bayesian framework can easily be extended to a hierarchical analysis of a cohort of patients with different individual characteristics (and state series) but with common structure in terms of heteroskedasticity parameters or state cross-correlations.  The relevance of such an analysis would be most pressingly felt in units where there was a degree of patient homogeneity, such as neo-natal intensive care, and be useful as it would allow common parameters to be more effectively inferred.

		\section*{Acknowledgements}
		A. M. Schmidt and D. A. Stephens are grateful for financial support from the Natural Sciences and Engineering Research Council (NSERC) of Canada (Discovery Grants).
		\vspace*{-8pt}

		\section*{Supporting information}
		Supporting \texttt{R} code and functions can be found in a \texttt{GitHub} repository linked below. All other information, descriptions and analysis can be found in Supplemental Materials document provided with online version of this manuscript.
		
		\noindent \texttt{GitHub: https://github.com/zaydomar29/MVT-GARCH-SSM.git}

		\bibliography{citations}

\begin{thebibliography}{10}
\providecommand \doibase [0]{http://dx.doi.org/}%

\bibitem{stein2013challenges}
Stein PK. Challenges of heart rate variability research in the ICU. {\it
  Critical Care Medicine} 2013\string; 41(2)\string: 666--667.

\bibitem{sow2010real}
Sow D, Biem A, Sun J, Hu J, Ebadollahi S. Real-time prognosis of ICU
  physiological data streams. {\it 2010 Annual International Conference of the
  IEEE Engineering in Medicine and Biology} 2010\string: 6785--6788.

\bibitem{saykrs1973analysis}
Saykrs B. Analysis of heart rate variability. {\it Ergonomics} 1973\string;
  16(1)\string: 17--32.

\bibitem{malik1998heart}
Malik M. Heart rate variability. {\it Current Opinion in Cardiology}
  1998\string; 13(1)\string: 36--44.

\bibitem{grogan2004reduced}
Grogan EL, Morris~Jr JA, Norris PR, et al. Reduced heart rate volatility: an
  early predictor of death in trauma patients. {\it Annals of Surgery}
  2004\string; 240(3)\string: 547.

\bibitem{acharya2006heart}
Acharya RU, Joseph KP, Kannathal N, Lim CM, Suri JS. Heart rate variability: a
  review. {\it Medical and Biological Engineering and Computing} 2006\string;
  44\string: 1031--1051.

\bibitem{baselli1987heart}
Baselli G, Cerutti S, Civardi S, et al. Heart rate variability signal
  processing: a quantitative approach as an aid to diagnosis in cardiovascular
  pathologies. {\it International Journal of Bio-Medical Computing}
  1987\string; 20(1-2)\string: 51--70.

\bibitem{jung2015implications}
Jung K, Shah NH. Heart rate variability: a review. {\it Journal of Biomedical
  Informatics} 2015\string; 58\string: 168--174.

\bibitem{montanoHR}
Montano N, Porta A, Cogliati C, et al. Heart rate variability explored in the
  frequency domain: a tool to investigate the link between heart and behavior.
  {\it Neuroscience \& Biobehavioral Reviews} 2009\string; 33(2)\string:
  71--80.

\bibitem{baselli1986spectral}
Baselli G, Cerutti S, Civardi S, et al. Spectral and cross-spectral analysis of
  heart rate and arterial blood pressure variability signals. {\it Computers
  and Biomedical Research} 1986\string; 19(6)\string: 520--534.

\bibitem{ghassemi2015multivariate}
Ghassemi M, Pimentel M, Naumann T, et al. A multivariate timeseries modeling
  approach to severity of illness assessment and forecasting in ICU with
  sparse, heterogeneous clinical data. {\it Proceedings of the AAAI conference
  on artificial intelligence} 2015\string; 29\string: 446--553.

\bibitem{li2016model}
Li-wei HL, Mark RG, Nemati S. A model-based machine learning approach to
  probing autonomic regulation from nonstationary vital-sign time series. {\it
  IEEE Journal of Biomedical and Health Informatics} 2016\string; 22\string:
  56--66.

\bibitem{alloghani2020prospects}
Alloghani M, Baker T, Al-Jumeily D, Hussain A, Mustafina J, Aljaaf AJ.
  Prospects of machine and deep learning in analysis of vital signs for the
  improvement of healthcare services. {\it Nature-Inspired Computation in Data
  Mining and Machine Learning} 2020\string: 113--136.

\bibitem{luo2016predicting}
Luo Y, Xin Y, Joshi R, Celi L, Szolovits P. Predicting ICU mortality risk by
  grouping temporal trends from a multivariate panel of physiologic
  measurements. {\it Thirtieth AAAI conference on artificial intelligence}
  2016\string; 30(1).

\bibitem{che2018recurrent}
Che Z, Purushotham S, Cho K, Sontag D, Liu Y. Recurrent neural networks for
  multivariate time series with missing values. {\it Scientific Reports}
  2018\string; 8(1)\string: 6085.

\bibitem{shillan2019use}
Shillan D, Sterne JA, Champneys A, Gibbison B. Use of machine learning to
  analyse routinely collected intensive care unit data: a systematic review.
  {\it Critical Care} 2019\string; 23\string: 1--11.

\bibitem{johnson2016machine}
Johnson AEW, Ghassemi MM, Nemati S, Niehaus KE, Clifton DA, Clifford GD.
  Machine learning and decision support in critical care. {\it Proceedings of
  the IEEE} 2016\string; 104(2)\string: 444--466.

\bibitem{sun2020review}
Sun C, Hong S, Song M, Li H. A review of deep learning methods for irregularly
  sampled medical time series data. {\it arXiv preprint arXiv:2010.12493} 2020.

\bibitem{johnson2017reproducibility}
Johnson AEW, Pollard TJ, Mark RG. Reproducibility in critical care: a mortality
  prediction case study. {\it Proceedings of the 2nd Machine Learning for
  Healthcare Conference} 2017\string: 361--376.

\bibitem{caballero2015dynamically}
Caballero~Barajas KL, Akella R. Dynamically modeling patient's health state
  from electronic medical records: A time series approach. {\it KDD'15:
  Proceedings of the 21th ACM SIGKDD International Conference on Knowledge
  Discovery and Data Mining} 2015\string: 69--78.

\bibitem{west2006bayesian}
West M, Harrison J. {\it Bayesian Forecasting and Dynamic Models}.
\newblock Springer Science \& Business Media .
\newblock 1997.

\bibitem{GARCH}
Bollerslev T. Generalized autoregressive conditional heteroskedasticity. {\it
  Journal of Econometrics} 1986\string; 31(3)\string: 307--327.

\bibitem{johnson2016db}
Johnson AEW, Pollard TJ, Mark R. MIMIC-III clinical database (version 1.4).
  {\it PhysioNet} 2016\string; 10(C2XW26)\string: 2.

\bibitem{johnson2016}
Johnson AEW, Pollard TJ, Shen L, et al. MIMIC-III, a freely accessible critical
  care database. {\it Scientific Data} 2016\string; 3(1)\string: 1--9.

\bibitem{aminikhanghahi2017survey}
Aminikhanghahi S, Cook DJ. A survey of methods for time series change point
  detection. {\it Knowledge and information systems} 2017\string; 51(2)\string:
  339--367.

\bibitem{adams2007bayesian}
Adams RP, MacKay DJC. Bayesian online changepoint detection. {\it arXiv
  preprint arXiv:0710.3742} 2007.

\bibitem{liu2020sequential}
Liu B, Qi Y, Chen KJ. Sequential online prediction in the presence of outliers
  and change points: an instant temporal structure learning approach. {\it
  Neurocomputing} 2020\string; 413\string: 240--258.

\bibitem{garnett2010sequential}
Garnett R, Osborne MA, Reece S, Rogers A, Roberts SJ. Sequential Bayesian
  prediction in the presence of changepoints and faults. {\it The Computer
  Journal} 2010\string; 53(9)\string: 1430--1446.

\bibitem{ruggieri2016exact}
Ruggieri E, Antonellis M. An exact approach to Bayesian sequential change point
  detection. {\it Computational Statistics \& Data Analysis} 2016\string;
  97\string: 71--86.

\bibitem{CCgarch}
Bollerslev T. Modelling the coherence in short-run nominal exchange rates: a
  multivariate generalized ARCH model. {\it The Review of Economics and
  Statistics} 1990\string: 498--505.

\bibitem{carter1994gibbs}
Carter CK, Kohn R. On Gibbs Sampling for State Space Models. {\it Biometrika}
  1994\string; 81(3)\string: 541--553.

\bibitem{fruhwirth1994data}
Fr{\"u}hwirth-Schnatter S. Data Augmentation and Dynamic Linear Models. {\it
  Journal of Time Series Analysis} 1994\string; 15(2)\string: 183--202.

\bibitem{cordoba2018fast}
C{\'o}rdoba I, Varando G, Bielza C, Larra{\~n}aga P. A fast Metropolis-Hastings
  method for generating random correlation matrices. {\it Intelligent Data
  Engineering and Automated Learning--IDEAL 2018: 19th International
  Conference, Madrid, Spain, November 21--23, 2018, Proceedings, Part I 19}
  2018\string: 117--124.

\bibitem{kitagawaIC}
Konishi S, Kitagawa G. {\it Information criteria and statistical modeling}.
\newblock Springer Science \& Business Media .
\newblock 2008.

\bibitem{gelman2013bayesian}
Gelman A, Stern HS, Carlin JB, Dunson DB, Vehtari A, Rubin DB. {\it Bayesian
  Data Analysis}.
\newblock Chapman and Hall/CRC .
\newblock 2013.

\bibitem{waic}
Watanabe S. Asymptotic equivalence of Bayes cross validation and widely
  applicable information criterion in singular learning theory. {\it Journal of
  Machine Learning Research} 2010\string; 11\string: 3571--3594.

\bibitem{Fruhwirth-Schnatter2006}
Fr{\"{u}}hwirth-Schnatter S. {\it Finite Mixture and Markov Switching Models}.
\newblock New York, NY: Springer New York .
\newblock 2006

\end{thebibliography}
		
		\newpage

		\section*{Appendix}
		\section{Simulation Study}
		
		We present a simulation study illustrating the performance of the model for a 4-d GARCH-SSM. We simulate a GARCH-SSM data using a sample size of $T=1000$ and we estimate the parameters using both the standard state space model (SSM) approach and our GARCH-SSM formulation.  The observation matrix, $\F$, and state evolution matrix, $\G$, are constant throughout time and fixed as $\F = \G = \bm{I}_4$, where $\bm{I}_4$ is the $4 \times 4$ identity matrix.  With the same parameter values we simulate 100 replicate data sets from a 4-dimensional GARCH(1,1)-SSM, thus allowing us to assess the frequentist behaviour in the parameter estimates.   The model parameters are the unknown components of the variance, that is the state variance and the components of the GARCH(1,1) error. We seek to estimate these parameters, the latent states and an estimate of the observation variance. The structure of $\F$ and $\G$ in this case represent a random-walk plus noise structure for the mean level.
		
		\subsection{Choice of priors}
		To estimate the model using the Bayesian framework we need to specify the priors. For the GARCH parameter $\alpha_0^{(i)}$ we choose a half-Cauchy prior; for $(\alpha_1^{(i)},\beta_1^{(i)})$ we choose a joint half-Cauchy prior restricted to the region such that $\alpha_1^{(i)}+\beta_1^{(i)}<1$; for the state covariance matrix, $\W$, we choose an Inverse-Wishart prior, $IW(10,10)$, where, $i,j=1,2,3,4$.  The choice of priors reflects the positivity restriction that we have made for the GARCH parameters  and the diagonal components of the correlation components. The half-Cauchy prior has a thicker tail than other priors such as the half-normal and this allows for larger values for $\alpha_0$ to occur more frequently. This has the effect of allowing the baseline variance for the GARCH process to be high. We have also made a stationarity restriction for the GARCH parameters as given by the support $\alpha_1^{(i)}+\beta_1^{(i)}<1$ for $i=1,2,3,4$. For the correlation parameters in the matrix $\bm{U}$ described below, we choose priors for the unnormalized version of $\bm{U}$ and then normalize the proposed quantity so that $\bm{R}=\bm{UU}^\top$ has unit diagonal entries. For the diagonal entries of the unnormalized version of $\bm{U}$ we choose a half-Cauchy prior, for the remainder of the correlation parameters, $u_{ij}$, we choose a $N(0,1)$ prior. Our choice of the Inverse-Wishart prior for the state variance $\W$, allows for a conjugate analysis. However, for all other parameters in the model we obtain samples from the posterior distributions using the Metropolis-Hastings algorithm.

		\subsection*{Model specification and Bias Estimation:}
		The 4d-GARCH(1,1)-SSM  is given by the following specification.
		\begin{align}
			\begin{split}\label{Model_spec}
				\text{Observation equation: } \qquad &\bm{y}_t = \F\bm{\theta}_t+ \bm{z}_t\\
				\text{State equation: } \qquad
				&\bm{\theta}_t = \G\bm{\theta}_{t-1}+\bm{w}_t, \qquad w_t\sim N(\bm{0},\bm{W})\\
				\text{GARCH error specification}: \qquad &
				\begin{cases}
					\bm{z}_t = \bm{V}^{1/2}_t\bm{\epsilon_t} , \qquad \bm{\epsilon_t} \sim N(\bm{0},\bm{I}_n)\\
					\bm{V}_t  = \bm{D}_t\bm{R}\bm{D}_t
				\end{cases}
			\end{split}
		\end{align}
		
		with $\bm{D_t} = \textrm{diag}(\sigma_{1,t},\dots,\sigma_{n,t})$ and, for $i=1,\dots,n$,
		\[
		\sigma_{i,t}^2 = \alpha_0^{(i)}+\alpha_1^{(i)}z_{i,t-1}^2+ \beta_1^{(i)}\sigma_{i,t-1}^2
		\]
		with $\alpha_{0}^{(i)},\alpha_{1}^{(i)},\beta_{1}^{(i)}\in [0,1]$, and
		\begin{align}
			\bm{R} &= \begin{bmatrix}
				1 & \rho_{1,2} & \dots & \rho_{1,n}\\
				\rho_{2,1} & 1 & \dots & \rho_{2,n}\\
				\vdots & & \ddots & \vdots\\
				\rho_{n,1} & \rho_{n,2} & \dots & 1\label{RTmatrix}
			\end{bmatrix} = \bm{U}\bm{U}^\top,
		\end{align}
		
		and the parametrization $\bm{U}$ represents the upper Cholesky decomposition of $\bm{R}$.
		
		\begin{align*}
			\begin{split}
				\bm{U}=
				\begin{pmatrix}
					u_{11} & u_{12} & \dots & u_{1n}\\
					0 & u_{22} & \dots & u_{2n}\\
					\vdots & \vdots & \ddots & \vdots \\
					0 & 0 &  \dots & u_{nn}\\
				\end{pmatrix}
			\end{split}
		\end{align*}

		In the simulation we used $\F=\G=\bm{I}_{n\times n}$ so that we have a random walk plus noise model. Below we provide the estimation results from the 100 simulations summarized in the form of bias. The bias is computed based on the posterior median estimates from each of the 100 simulations. First, reported are the bias of the GARCH parameters; next, we report the bias for the estimates of the unknown state Variance matrix, $\bm{W}$; finally, we report the boxplot based on the posterior median estimates for the parameters of $\bm{U}$.

		\begin{table}[H]
			\centering
			\def\~{\hphantom{0}}
			
			\begin{tabular*}{\columnwidth}{@{\extracolsep{\fill}}l@{\extracolsep{\fill}}l@{\extracolsep{\fill}}r@{\extracolsep{\fill}}r@{}r}
				
				Series & & $\bm{\alpha_{0}}$ & $\bm{\alpha_{1}}$ & $\bm{\beta_{1}}$ \\ 
				\hline
				\multirow{3}{*}{$\bm{Y_1}$}&
				\textbf{True} & $\bm{1.00}$ & $\bm{0.10}$ &$\bm{0.80}$\\ 
				\cline{2-5}
				& Bias & 0.18 & 0.03 & -0.04 \\ 
				& Std. Error  & (0.17) & (0.01) & (0.02) \\
				\hline
				\multirow{3}{*}{$\bm{Y_2}$}&
				\textbf{True} & $\bm{1.00}$ & $\bm{0.30}$ & $\bm{0.60}$ \\
				\cline{2-5} 
				& Bias  & 0.29 & 0.12 & -0.15 \\
				& Std. Error  & (0.17) & (0.03) & (0.03) \\
				\hline
				\multirow{3}{*}{$\bm{Y_3}$}&
				\textbf{True} & $\bm{2.00}$ &$\bm{ 0.10}$ & $\bm{0.40}$ \\ 
				\cline{2-5}
				& Bias & 0.51 & 0.04 & -0.17 \\
				& Std. Error & (0.45) & (0.02) & (0.11) \\ 
				\hline
				\multirow{3}{*}{$\bm{Y_4}$}&
				\textbf{True} & $\bm{2.00}$ & $\bm{0.20$} & $\bm{0.70}$ \\
				\cline{2-5} 
				&Bias & 0.55 & 0.07 & -0.09 \\
				&Std. Error & (0.28) & (0.02) & (0.03) \\
				\hline 
			\end{tabular*}\vskip18pt 
			\caption{GARCH Parameter Estimates and Biases} \label{tableone}
		\end{table}
		
		\begin{align}
			\begin{split}
				Bias(\bm{W}) = 
				\begin{pmatrix}
					\bm{-0.028 }  & \bm{0.000} &\bm{0.007} &  \bm{0.008} \\ (0.014 )& (0.004 ) & (0.004 ) & (0.003)\\
					& \bm{0.005} & \bm{0.000  } &\bm{0.001}\\-& (0.009)&(0.004)&(0.003) \\
					-& -& \bm{0.014} &\bm{0.005}\\-& -& (0.009)&(0.005) \\
					-& -& -& \bm{0.020}\\-& -& -& (0.009)\\
				\end{pmatrix}\quad  
			\end{split} \label{eq: bias_W}
		\end{align}
		
		The true value of $\W$ used to generate the model was $\bm{W}=diag(0.1,0.1,0.1,0.1)$. From (\ref{eq: bias_W}) we can see that with respect to estimating the unknown state variance $\bm{W}$ the model is doing well. Next, for the correlation coefficient matrix, $\bm{U}$, we see from the boxplots in Figure \ref{fig: boxplot_corr} that the model is also able to estimate this parameter well. Table \ref{tableone} shows that the GARCH parameters are being estimated reasonably well. Overall owing to the latent structure and the multidimensionality of the model it becomes difficult to estimate the GARCH parameters. It must be noted again that the latent state also  must be estimated in order to calculate the GARCH paramaters.
		
		Finally, it still remains to be checked that for each of these simulated data, the GARCH-SSM is being selected as the most appropriate model based on our WAIC selection criteria. Figure \ref{fig: boxplotWAIC} shows the boxplots of the WAIC estimated when each of the simulated data sets are estimated using a GARCH-SSM versus a standard-SSM. We see that the WAIC is correctly selecting the GARCH-SSM over the standard-SSM. This can be understood from the fact that the WAIC of the GARCH model is greater than that of the standard-SSM. The panel on the right in Figure \ref{fig: boxplotWAIC} is a histogram of the difference in the GARCH and the standard-SSM based WAICs computed for each data set. This also reinforces the finding that the WAIC is correctly selecting the appropriate model, since for almost all of the simulated data sets we have that the difference in the WAICs is positive, thus showing that the GARCH based WAIC is greater than the standard SSM based WAIC.
		
		\begin{figure}[H]
			\begin{center}
				\includegraphics*[width=7cm, height=7cm]{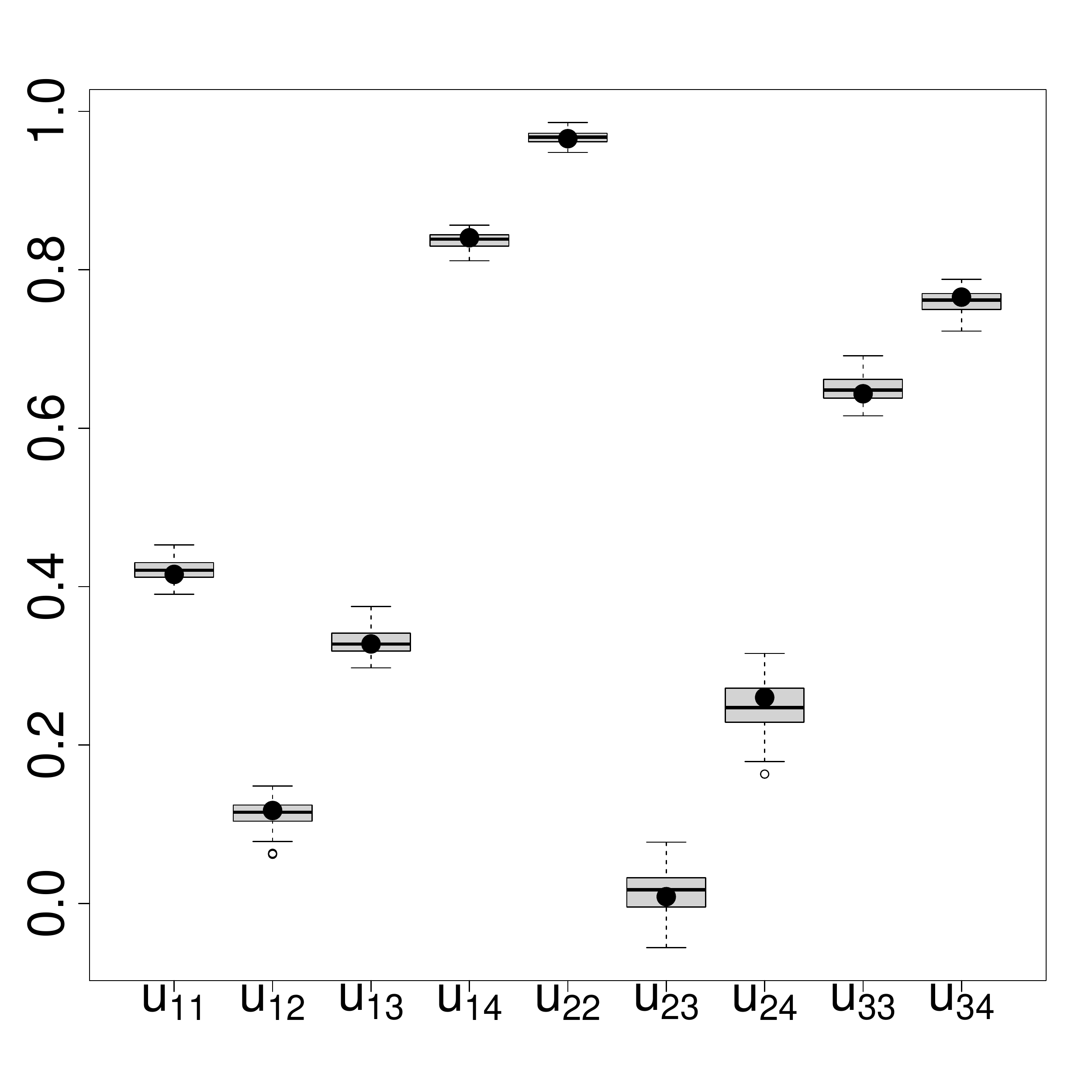}
				\caption{Boxplot: Distribution of the correlation components of $\bm{U}$ based on posterior median estimates of the 100 simulated data sets. The bold black solid circles represent the true values. \label{fig: boxplot_corr}}	
			\end{center}
		\end{figure}
		
		\begin{figure}[H]
			
			\begin{center}
				\includegraphics*[width=6cm, height=7cm]{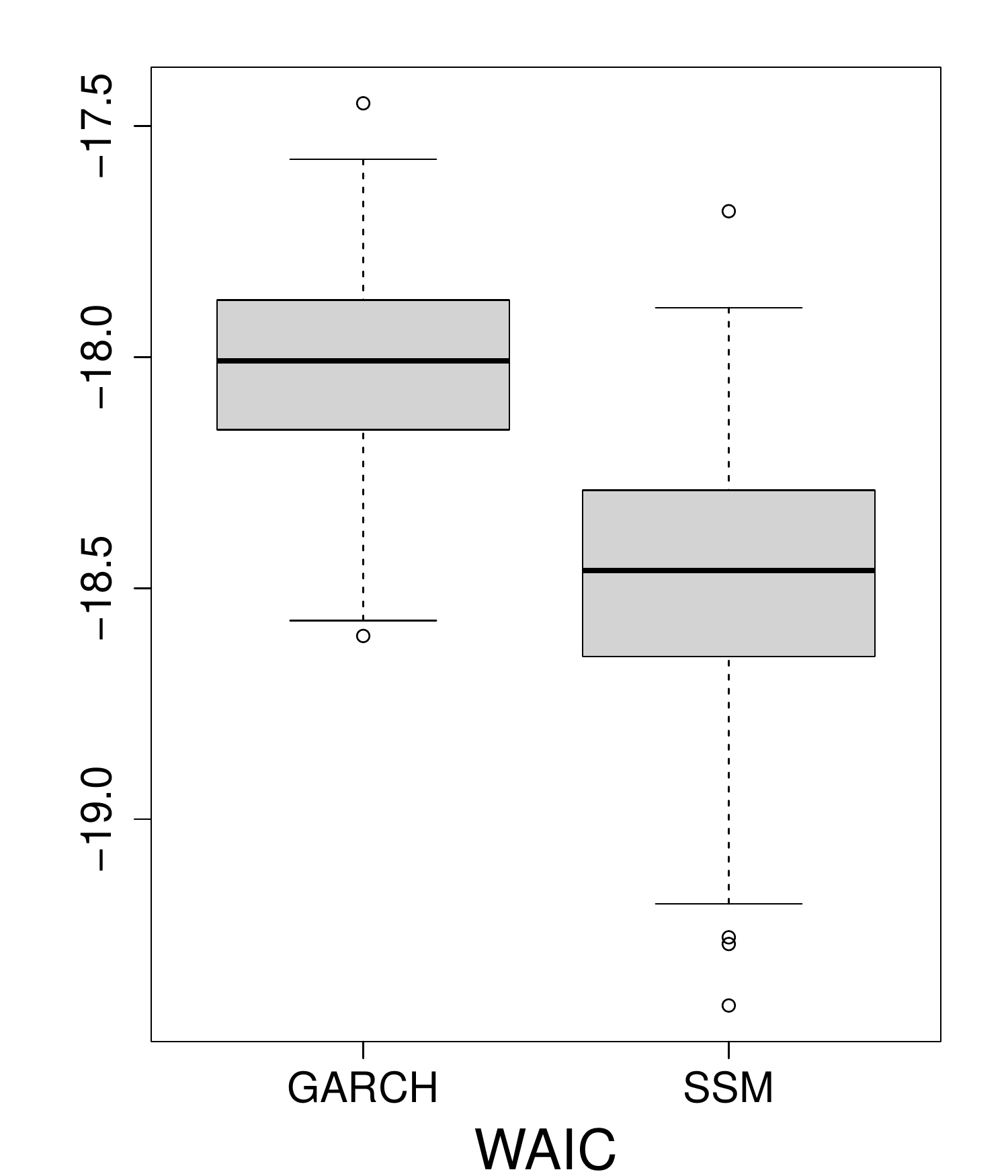}
				\includegraphics*[width=8cm, height=7cm]{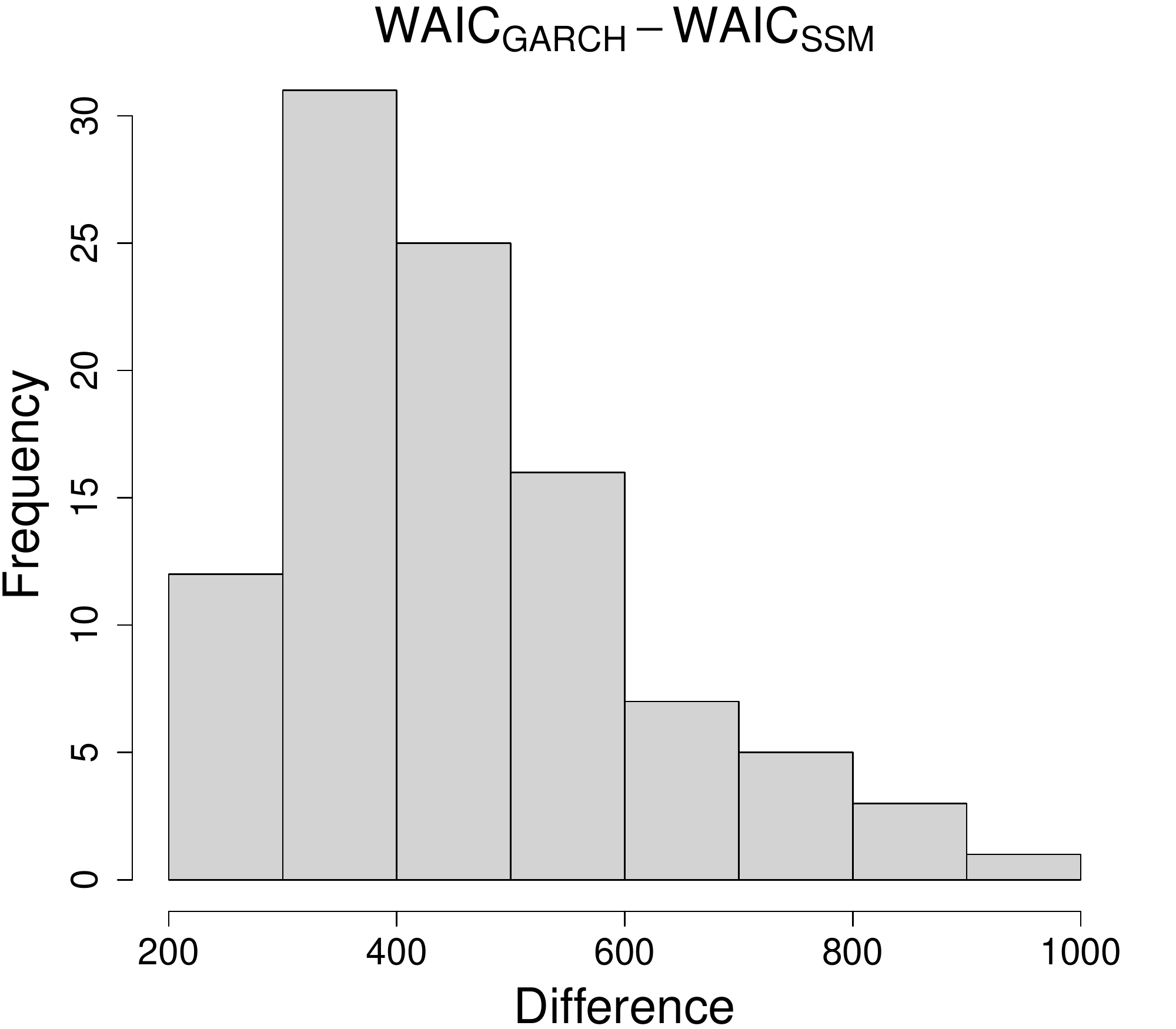}
			\end{center}
			\caption{Left: The boxplot shows the comparison of the WAIC estimates of the 100 simulated data when each data is estimated using a GARCH-SSM versus a standard-SSM. Y-axis is scaled down by a factor of a 1000. Right: A histogram of the difference between the WAIC from the GARCH estimates and the standard SSM. \label{fig: boxplotWAIC}}	
		\end{figure}

		\newpage
		
		\section{Model Estimation and Residual Analysis}
		\subsection{Model Estimation}
		For illustration, the following results are for a single 4-dimensional series of length $T=1000$. Figure \ref{OneStepFore} shows the posterior estimates of the one-step-ahead forecast, $\bm{f}_t$. We can see that this basically traces out the mean level of the observational series for each series. Figure \ref{PostState} shows the posterior estimates of the latent state for each of the series in thick black and the true state vector in the broken lines. We can see that we are able to estimate and recover the unobserved states accurately and the true value of the unobserved state almost always lies in the 95\% credible interval. In the data analysis of real data, when we apply this model to the HR and BP data, the posterior estimates of the latent state will allow us to study the variability of the HR and BP over the time period of the data.

		The panels in Figure \ref{DynVar} shows the posterior estimates of the standard deviation in black and in gray we have the true standard deviation. This plot shows the flexibility our proposed model provides over the standard Gaussian state-space model. We know that for the data the variance is not constant and being able to estimate it properly allows us to draw better inference on the process at a particular time, $t$. We can see that our estimate for the dynamic standard deviation is quite close to the actual standard deviation of the observed series and the true values are recovered.

		The 95\% credible intervals of both the estimated state vector and the variances contains the true value of the state vector and the dynamic variances. This shows that our model is able to accurately capture both latent state and the unknown dynamic variance.

		\subsection{Residual Analysis}
		Next, we carry out a residual analysis. Recall that by assumption from (\ref{Model_spec}) that the GARCH errors when normalized with respect to the dynamic variance should result in a standard normal error distribution. Replicating this result empirically would be evidence that our model is able to capture the correct heteroskedastic  structure of the errors. From Figure (\ref{DynVar}) we can see that for the second series, there is quite a bit of heteroskedasticity. Thus one would suspect that compared to the quantiles of the standard normal distribution, this series would show large deviations. And this is exactly what can be seen in Figure \ref{unnorm_var_qq} (top right panel). Consequently, normalizing all the series with their true heterskedasticity should result in normalized errors that adhere strongly to the quantiles of the standard normal distribution. This second result can be seen in Figure \ref{norm_var_qq}.
		
		\begin{figure}[H]
			\begin{center}
				\includegraphics[width = 1.0\textwidth,height=0.5\textwidth]{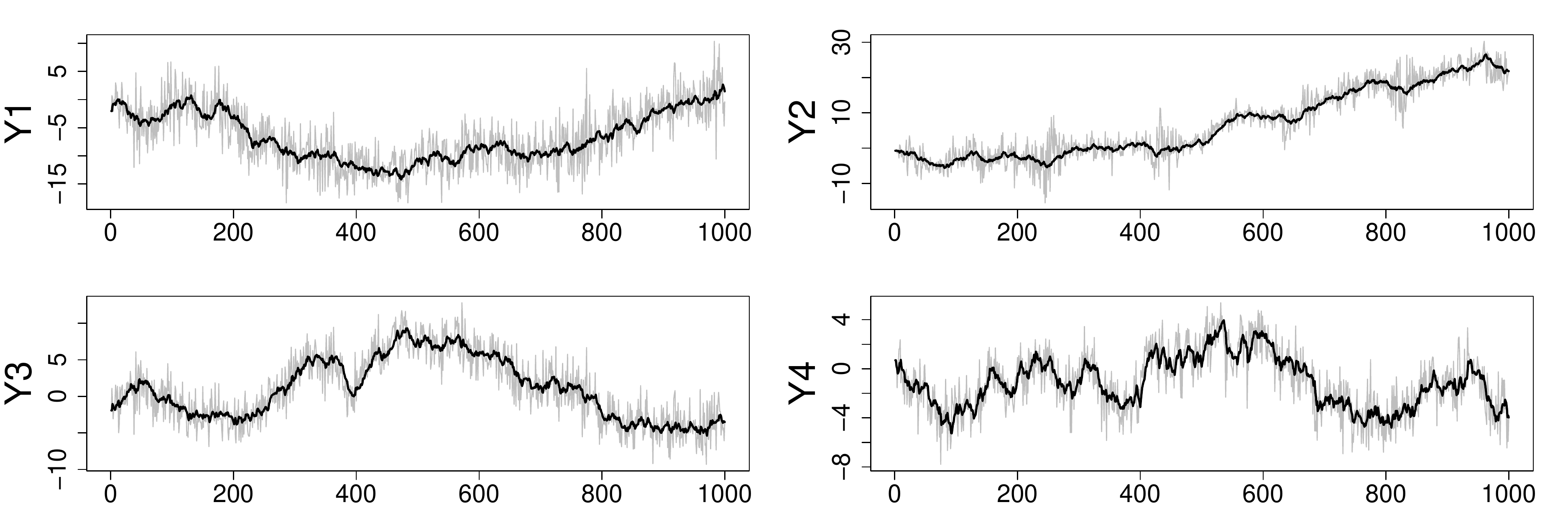}
			\end{center}
			\caption{Simulated data: the posterior estimates of the one step ahead forecasts of each series. In gray are the observed data.}\label{OneStepFore}
			
			\begin{center}
				\includegraphics[width = 1.0\textwidth,height=0.3\textwidth]{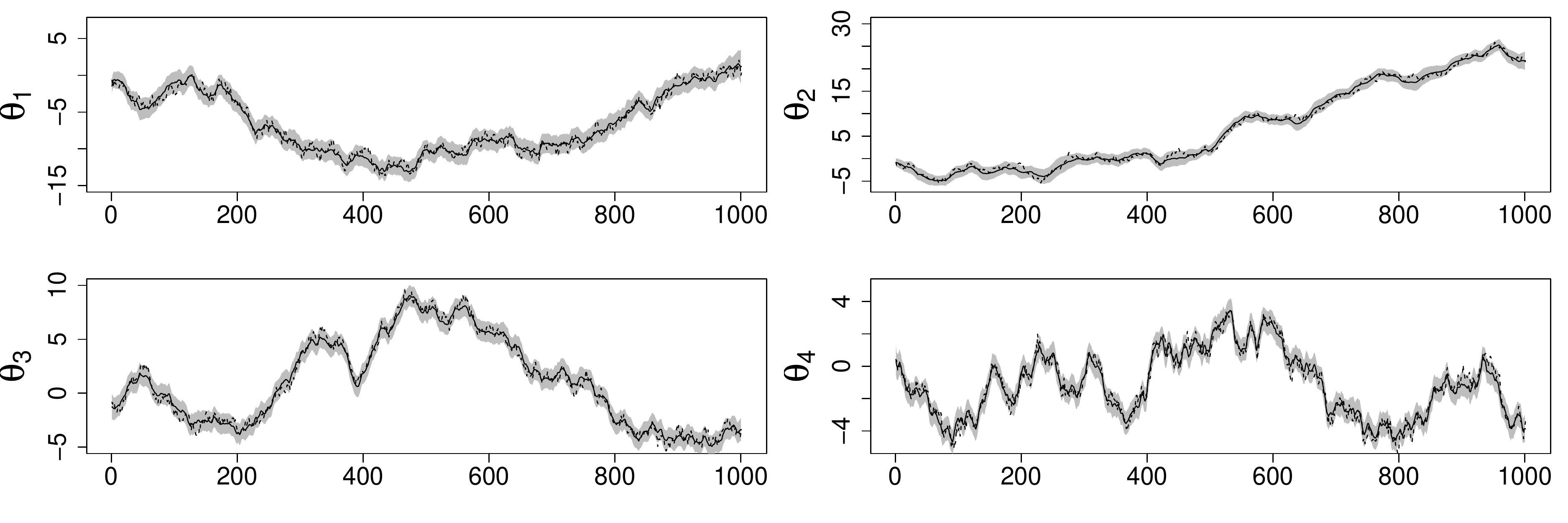}
			\end{center}
			\caption{Simulated data:  The thick dark line are the posterior median estimates of the states. The dashed lines are the true value of the states. The gray band is the 95\% posterior credible interval.}\label{PostState}
			\begin{center}
				\includegraphics[width = 1.0\textwidth,height=0.35\textwidth]{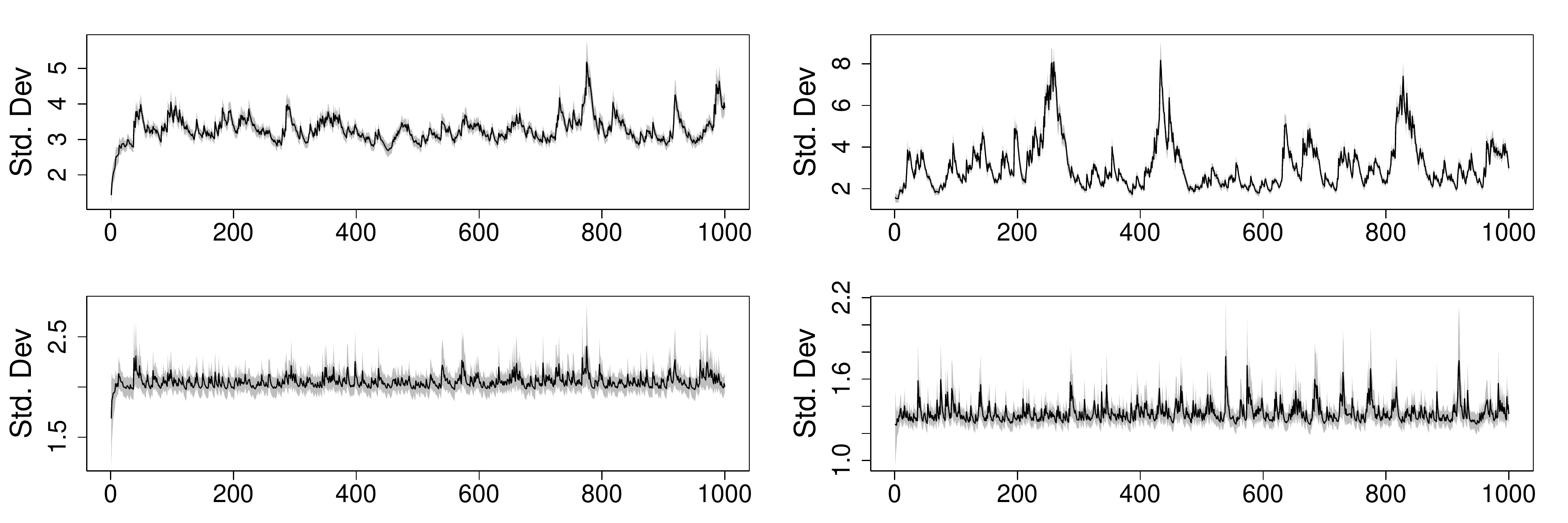}
			\end{center}
			\vspace{0.5cm}\caption{Simulated data: The true dynamic standard deviation in black. The shaded gray area is the 95\% posterior credible interval of the estimated standard deviation.}\label{DynVar}
			
		\end{figure}

		\begin{figure}[H]
			\begin{center}
				\includegraphics[width = 0.9\textwidth, height=0.45\textwidth]{ 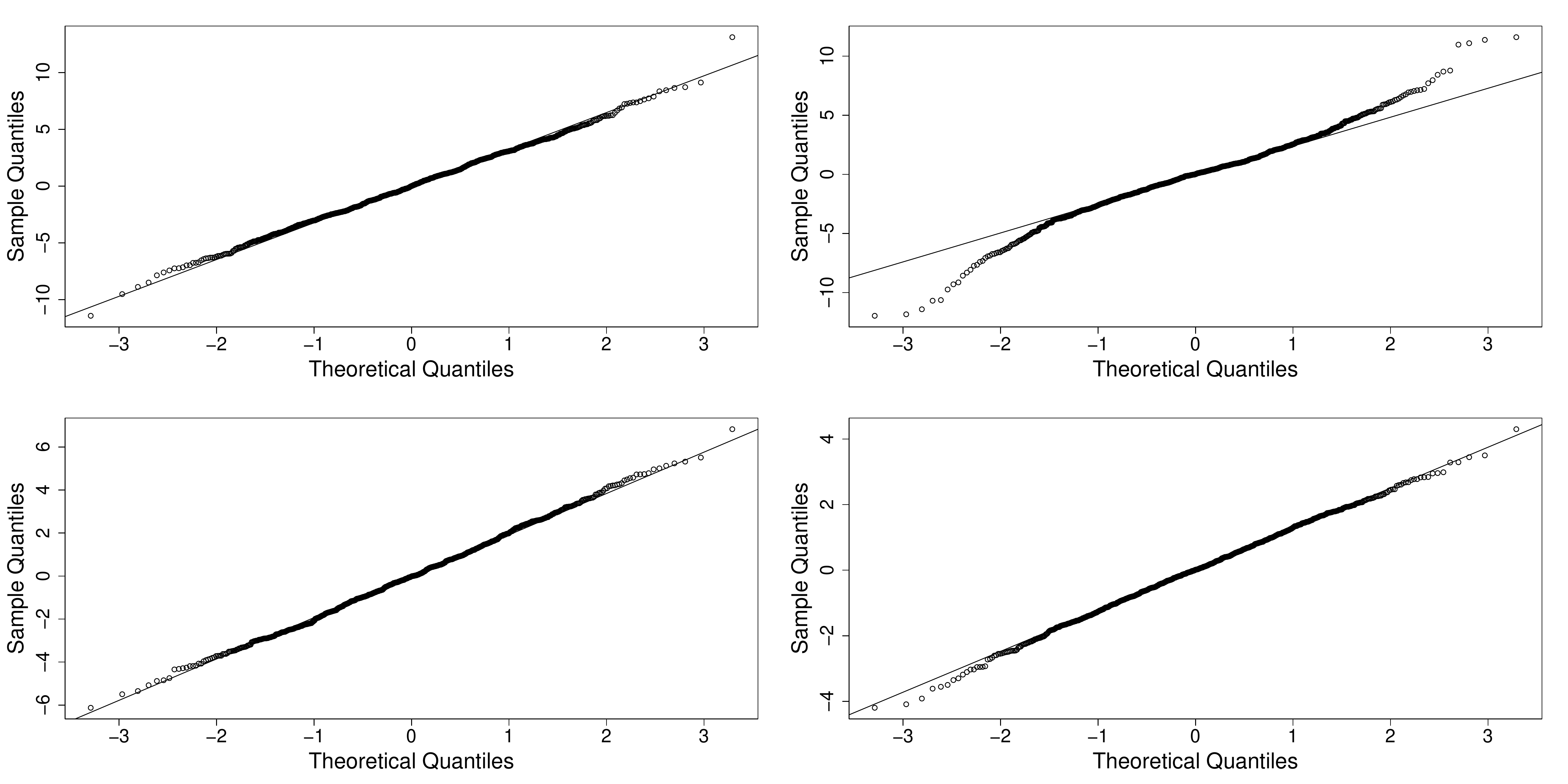}
			\end{center}
			\vspace{0.5cm}\caption{Normal QQ-plots for the un-normalized estimated error terms. Top (left to right): Y1 and Y2. Bottom row (left to right): Y3 and Y4.}\label{unnorm_var_qq}
			\begin{center}
				\includegraphics[width = 0.9\textwidth, height=0.45\textwidth]{ 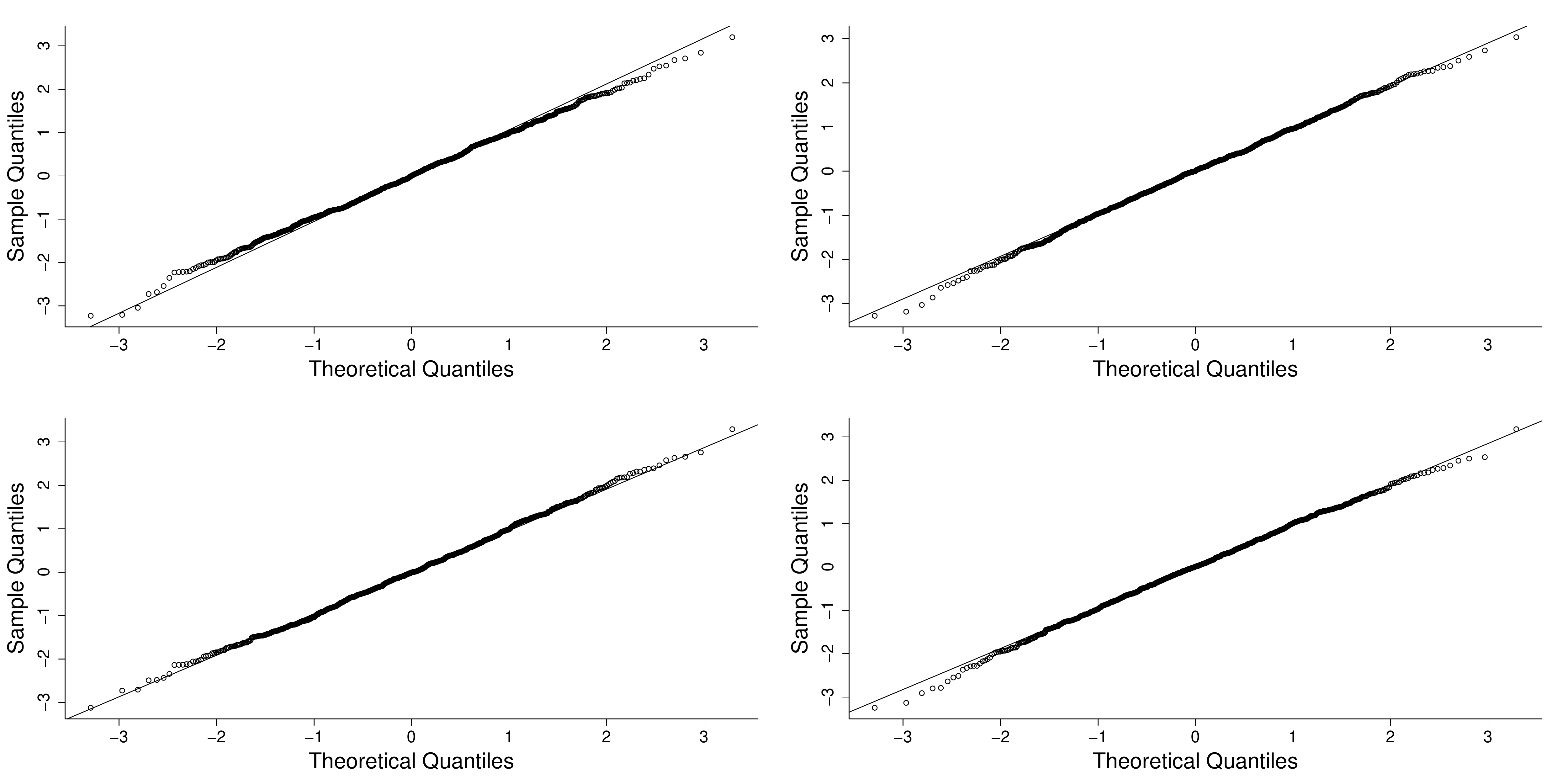}
			\end{center}
			\vspace{0.5cm}\caption{Heteroskedasdicity adjusted normal QQ-plots for the estimated error terms. Top (left to right): Y1 and Y2. Bottom (left to right): Y3 and Y4.}\label{norm_var_qq}
		\end{figure}

		\newpage
		\section{Secondary Analysis}
		
		In this section we analyze data from some more patient data from the MUHC data set. Below we have the estimates for a second patient using the GARCH-model and the following that the estimates using a standard SSM. We see that in this case, the standard SSM is more appropriate than the the GARCH-model when using the WAIC.

		\begin{table}[H]
			\centering
			\def\~{\hphantom{0}}
			\label{t:tablefour}
			\begin{tabular*}{\columnwidth}{@{\extracolsep{\fill}}c@{\extracolsep{\fill}}c@{\extracolsep{\fill}}c@{\extracolsep{\fill}}c@{}c}
				% \hline
				% \rowcolor{lightgray}\multicolumn{5}{c}{\underline{GARCH(1,1)-SSM Model}} \\
				\multicolumn{5}{c}{\underline{Estimates for HR series}} \\
				& $\bm{\alpha}^{HR}_{0}$ & $\bm{\alpha}^{HR}_{1}$& $\bm{\beta}^{HR}_{1}$& $\bm{W}^{HR}$\\
				\hline
				\textbf{Post. est.} & 6.321 & 0.376 & 0.429 &  4.150\\
				\textit{s.d.}& (2.278) & (0.093) & (0.122) & (0.754) \\
				CI. 95\%& (2.95, 11.71) & (0.23, 0.58) & (0.18, 0.65) & (2.97,5.88)  \\[12pt]
				\hline
				\multicolumn{5}{c}{\underline{Estimates for BP series}} \\[6pt]
				& $\bm{\alpha}^{BP}_{0}$ & $\bm{\alpha}^{BP}_{1}$& $\bm{\beta}^{BP}_{1}$& $\bm{W}^{BP}$ \\	
				\hline
				\textbf{Post. est.} & 9.574 & 0.278 & 0.163 & 0.923 \\
				\text{s.d.}& (2.162) & (0.088) & (0.136) & (0.227)\\
				CI. 95\%& (4.92, 13.31) & (0.14, 0.48) & (0.01, 0.52) & (0.58, 1.47) \\[6pt]
				\hline
				\multicolumn{5}{c}{\underline{Estimates of Correlation Parameters}} \\
				& $\bm{\rho}$ & $\bm{\rho}_s$ & & \\ 
				\hline
				\textbf{Post. est.} & 0.0197 & 0.251 && \\
				\text{s.d.}&(0.017) & (0.923) & & \\
				CI 95\%  & $(0.0008,  0.065)$ & $(-0.026, 0.59)$ & & \\
				& & & & \\
				\hline
				\textbf{WAIC} & -5508 & & \\
				\hline
			\end{tabular*}\vskip6pt
			\caption{Montreal ICU data:  Bivariate GARCH(1,1)-SSM parameter estimates. $\rho$ is the estimate of the correlation between the observation errors of the HRT series and the BP series and $\rho_s$ is the estimate of the correlation between the state errors of both series. The estimates are the posterior mean values and in the parenthes is below is their standard error and below that the $95\%$ credible interval.}\label{GarchEstHRBP}
		\end{table}

		\begin{table}[H]
			\centering
			\def\~{\hphantom{0}}
			\begin{tabular*}{\columnwidth}{@{\extracolsep{\fill}}c@{\extracolsep{\fill}}l@{\extracolsep{\fill}}c@{\extracolsep{\fill}}c}
				% \hline	
				% \rowcolor{lightgray}\multicolumn{4}{c}{\underline{Standard SSM Model}}\\
				\multicolumn{4}{c}{\underline{Observation Level Parameter Estimates}} \\
				& $\bm{V}_{HR}$ & $\bm{V}_{BP}$& $\bm{\rho}_{obs}$\\ % \cmidrule{1-4}
				\hline
				\textbf{Post. est.} & 13.578  &  12.684  &  0.074  \\
				\text{s.d.} & (3.38)  &  (1.381)  &  (1.092)\\
				\hline
				\multicolumn{4}{c}{\underline{State Level Parameter Estimates}} \\
				& $\bm{W}_{HR}$ & $\bm{W}_{BP}$& $\bm{\rho}_s$ \\ %\cmidrule{1-4}
				\hline
				\textbf{Post. est.} &  9.927  &  2.177  &  0.225\\
				\text{s.d.} & (4.879)  &  (0.288)  &  (0.333)\\ 
				\hline
				& & & \\
				\textbf{WAIC} & -5483 & & \\
				\hline
			\end{tabular*}\vskip6pt
			\vspace{0.2cm}\caption{Montreal ICU data: Parameter estimates of the bivariate Gaussian-SSM. Here $\rho_{obs}$ is the estimate of the correlation between the observation errors of the HRT series and the BP series and $\rho_s$ is the estimate of the correlation between the state errors of both series. The values displayed are the posterior means of the parameters and the values in the parenthesis are the standard errors.}\label{SSMestHRBP}
		\end{table}

		\begin{figure}[H]
			\begin{center}
				\includegraphics[width = 1\textwidth]{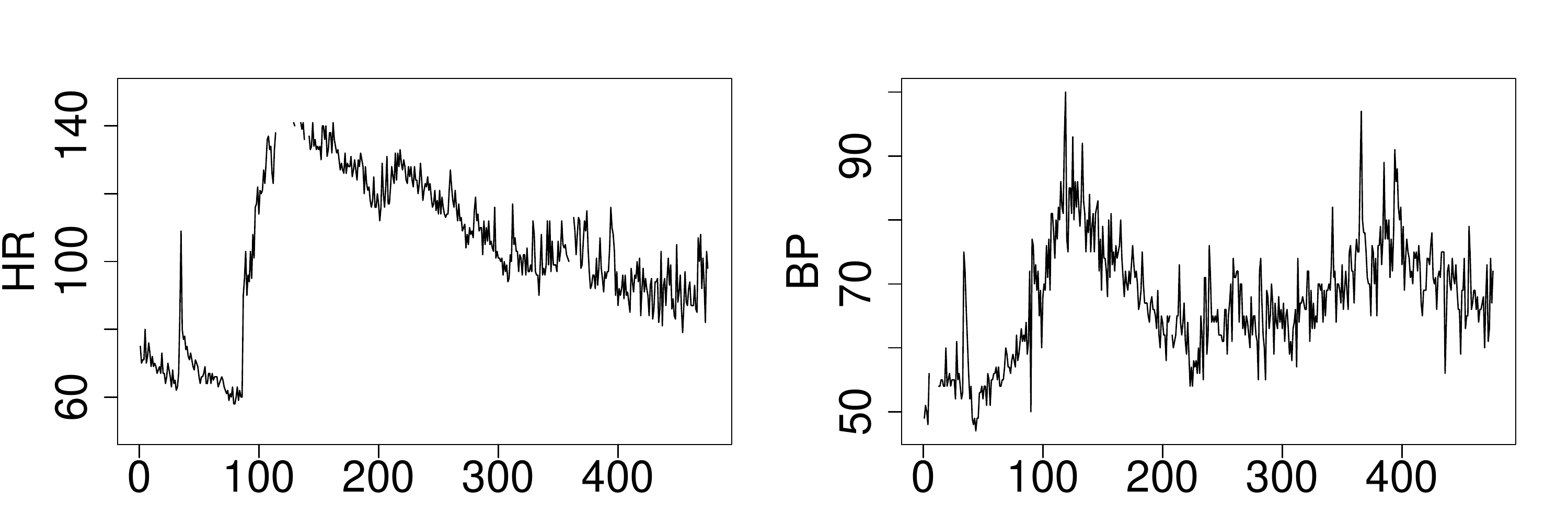}
			\end{center}
			\caption{Left: Heart rate series of a patient; Right: Blood pressure series of same patient. Time axis is number of seconds since start of monitoring\label{HrBpPlot3}}
			
			\begin{center}
				\includegraphics[width=1\textwidth,height=0.3\textwidth]{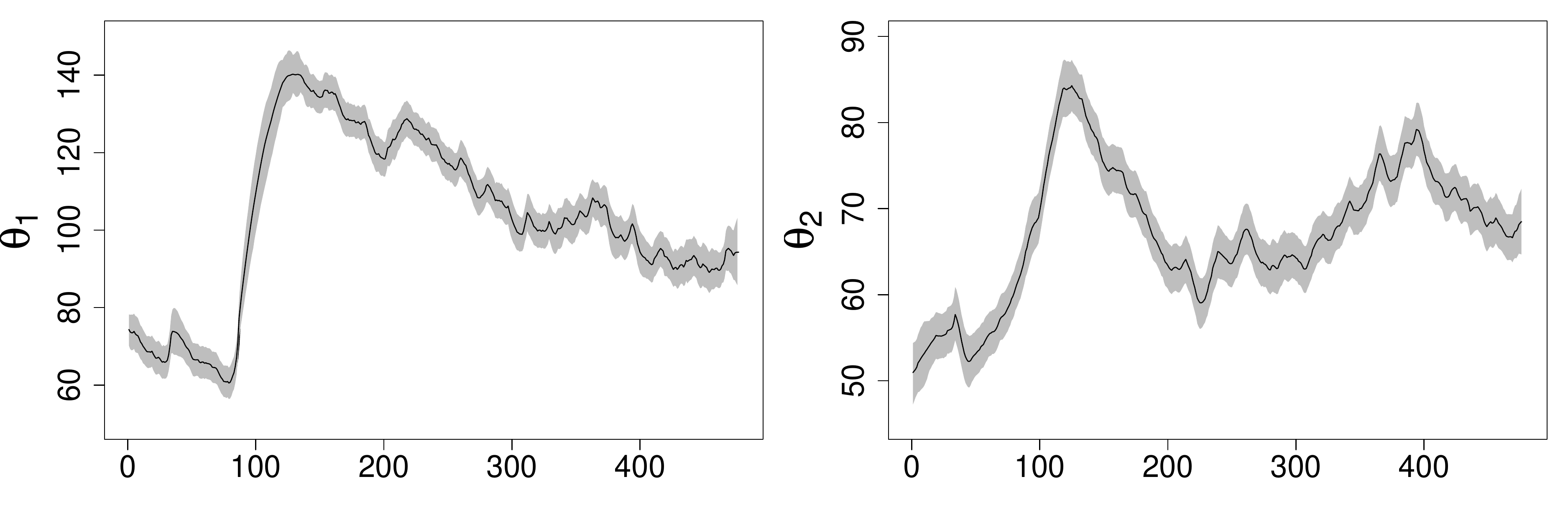}
			\end{center}
			\caption{Montreal ICU data.  Top row: posterior estimates of the mean level of each series with the credible intervals given by the dashed lines and the observed data in gray (HR left, BP right).}
			\begin{center}
				\includegraphics[width=1\textwidth,height=0.3\textwidth]{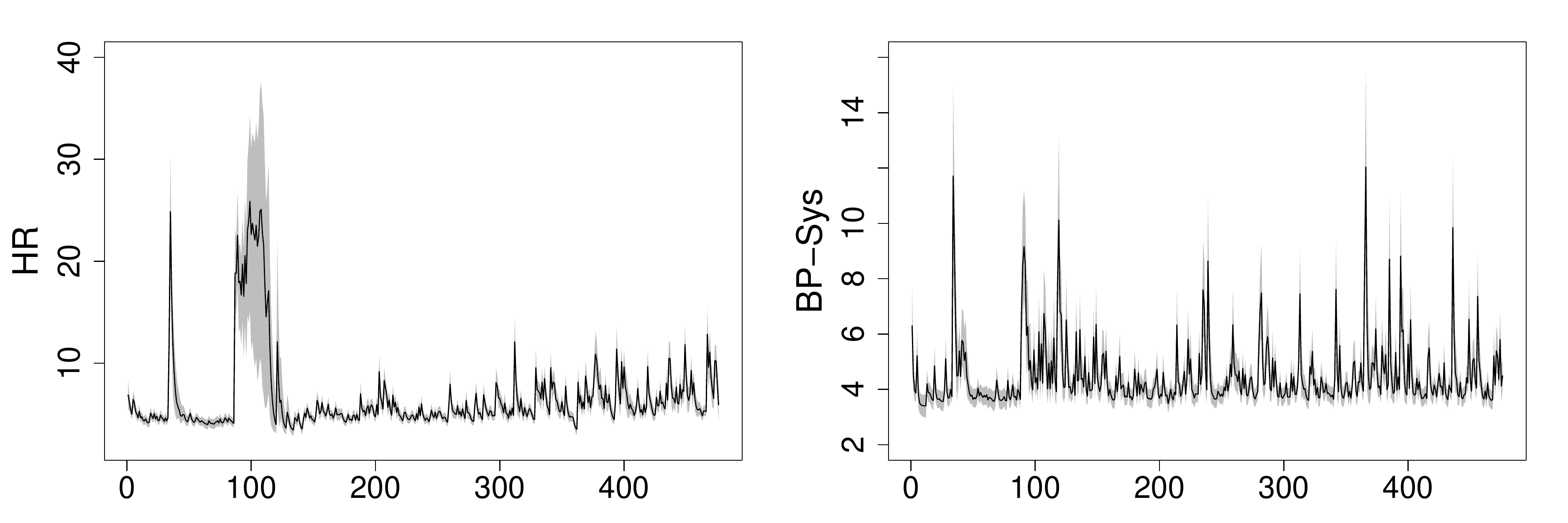}
			\end{center}
			\caption{Posterior mean estimates of the dynamic standard deviation (HR left, BP right).}
		\end{figure}

		\newpage
		The following is an analysis from a third patient. The estimates and plots are provided below. In this case, based on the WAIC we see that the GARCH model is the most appropriate.
		\begin{table}[H]
			\centering
			\def\~{\hphantom{0}}
			\label{t:tablefour}
			\begin{tabular*}{\columnwidth}{@{\extracolsep{\fill}}c@{\extracolsep{\fill}}c@{\extracolsep{\fill}}c@{\extracolsep{\fill}}c@{}c}
				% \hline
				% \rowcolor{lightgray}\multicolumn{5}{c}{\underline{GARCH(1,1)-SSM Model}} \\
				\multicolumn{5}{c}{\underline{Estimates for HR series}} \\
				& $\bm{\alpha}^{HR}_{0}$ & $\bm{\alpha}^{HR}_{1}$& $\bm{\beta}^{HR}_{1}$& $\bm{W}^{HR}$\\
				\hline
				\textbf{Post. est.} & 0.181 & 0.192 & 0.789 &  0.327\\
				\text{s.d.}& (0.044) & (0.031) & (0.031) & (0.048)\\
				CI. 95\%& (0.11, 0.28) & (0.14, 0.26) & (0.72, 0.84) & (0.246,0.428)  \\[12pt]
				\hline
				\multicolumn{5}{c}{\underline{Estimates for BP series}} \\[6pt]
				& $\bm{\alpha}^{BP}_{0}$ & $\bm{\alpha}^{BP}_{1}$& $\bm{\beta}^{BP}_{1}$& $\bm{W}^{BP}$ \\	
				\hline
				\textbf{Post. est.} & 0.700 & 0.456 & 0.510 &  0.388\\
				\text{s.d.}& (0.142) & (0.055) & (0.053) & (0.051)\\
				CI. 95\%& (0.47, 1.03) & (0.35, 0.56) & (0.41, 0.62) & (0.30,0.50) \\[6pt]
				\hline
				\multicolumn{5}{c}{\underline{Estimates of Correlation Parameters}} \\
				& $\bm{\rho}$ & $\bm{\rho}_s$ & & \\ 
				\hline
				\textbf{Post. est.} & 0.0512 & 0.031 && \\
				\text{s.d.}&(0.012) & (0.046) & & \\
				CI 95\%  & $(0.0295,  0.078)$ & $(-0.081, 0.10)$ & & \\
				& & & & \\
				\hline
				\textbf{WAIC} & -20434 & & & \\
				\hline
			\end{tabular*}\vskip6pt
			\caption{Montreal ICU data:  Bivariate GARCH(1,1)-SSM parameter estimates. $\rho$ is the estimate of the correlation between the observation errors of the HRT series and the BP series and $\rho_s$ is the estimate of the correlation between the state errors of both series. The estimates are the posterior mean values and in the parenthes is below is their standard error and below that the $95\%$ credible interval.}\label{GarchEstHRBP2}
		\end{table}

		\begin{table}[H]
			\centering
			\def\~{\hphantom{0}}
			\begin{tabular*}{\columnwidth}{@{\extracolsep{\fill}}c@{\extracolsep{\fill}}l@{\extracolsep{\fill}}c@{\extracolsep{\fill}}c}
				% \hline	
				% \rowcolor{lightgray}\multicolumn{4}{c}{\underline{Standard SSM Model}}\\
				\multicolumn{4}{c}{\underline{Observation Level Parameter Estimates}} \\
				& $\bm{V}_{HR}$ & $\bm{V}_{BP}$& $\bm{\rho}_{obs}$\\ % \cmidrule{1-4}
				\hline
				\textbf{Post. est.} & 8.94  &  3.611  &  0.050  \\
				\text{s.d.} & (0.367)  &  (0.249)  &  (0.342)\\
				\hline
				\multicolumn{4}{c}{\underline{State Level Parameter Estimates}} \\
				& $\bm{W}_{HR}$ & $\bm{W}_{BP}$& $\bm{\rho}_s$ \\ %\cmidrule{1-4}
				\hline
				\textbf{Post. est.} &  1.323  &  3.456  &  0.062\\
				\text{s.d.} & (0.196)  &  (0.314)  &  (0.532)\\ 
				\hline
				& & & \\
				\textbf{WAIC} & -22792 & &\\
				\hline
			\end{tabular*}\vskip6pt
			\vspace{0.2cm}\caption{Montreal ICU data: Parameter estimates of the bivariate Gaussian-SSM. Here $\rho_{obs}$ is the estimate of the correlation between the observation errors of the HRT series and the BP series and $\rho_s$ is the estimate of the correlation between the state errors of both series. The values displayed are the posterior means of the parameters and the values in the parenthesis are the standard errors.}\label{SSMestHRBP2}
		\end{table}

		\begin{figure}[H]
			\begin{center}
				\includegraphics[width = 1\textwidth]{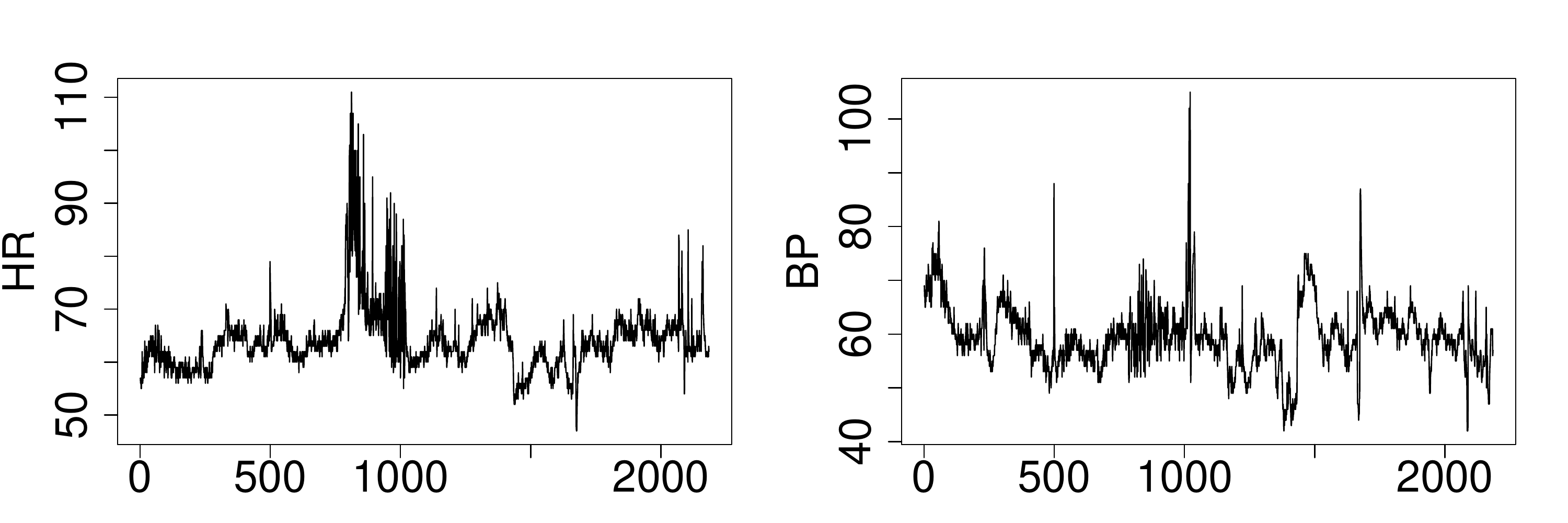}
			\end{center}
			\caption{Left: Heart rate series of a patient; Right: Blood pressure series of same patient. Time axis is number of seconds since start of monitoring\label{HrBpPlot2}}
			
			\begin{center}
				\includegraphics[width=1\textwidth,height=0.3\textwidth]{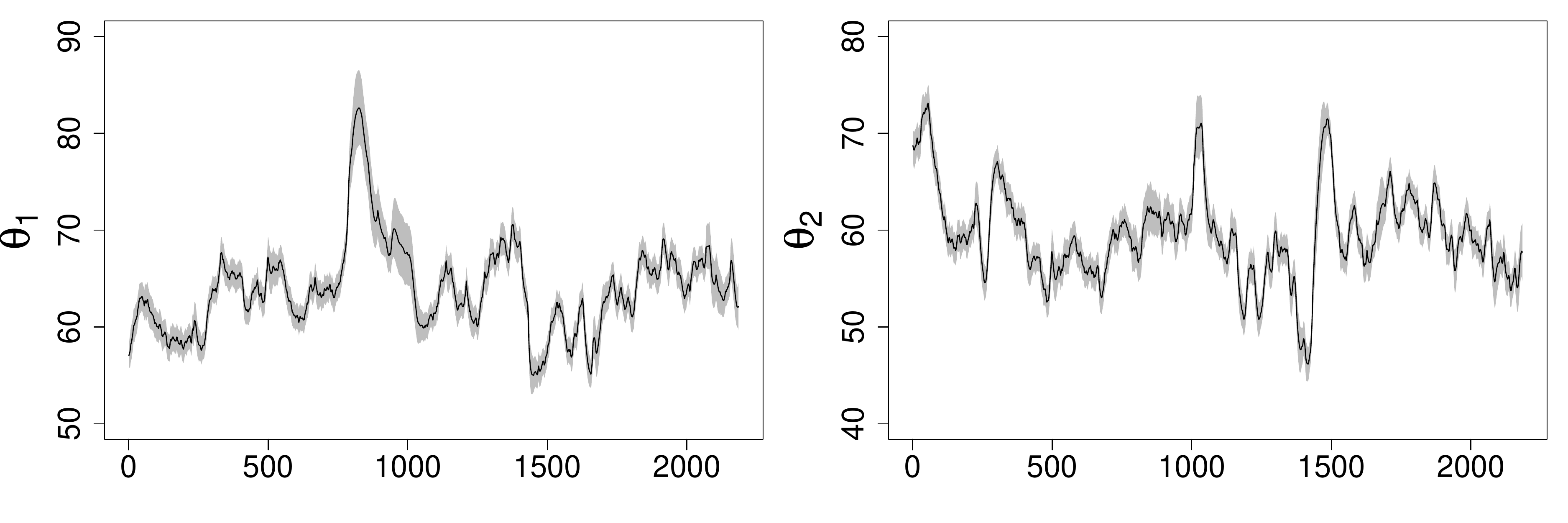}
			\end{center}
			\caption{Montreal ICU data.  Top row: posterior estimates of the mean level of each series with the credible intervals given by the dashed lines and the observed data in gray (HR left, BP right).}
			\begin{center}
				\includegraphics[width=1\textwidth,height=0.3\textwidth]{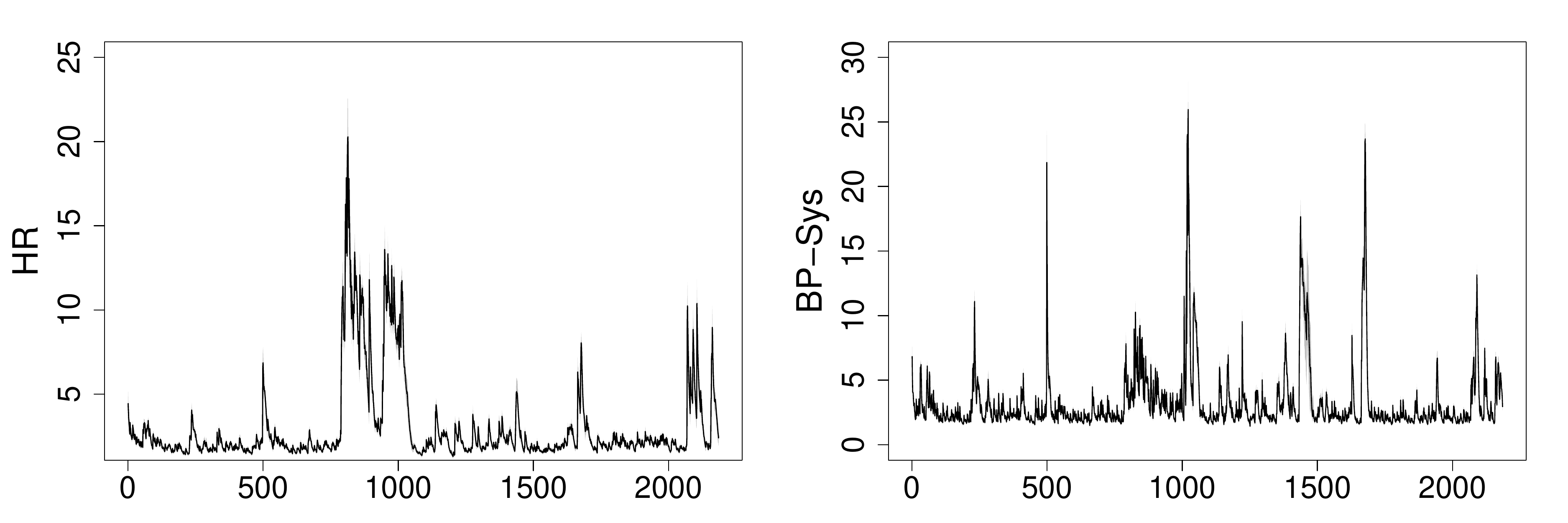}
			\end{center}
			\caption{Posterior mean estimates of the dynamic standard deviation (HR left, BP right).}
		\end{figure}

		\section{Calculation of the WAIC}
		The $lppd$ is given by,
		\begin{align}
			\begin{split}
				lppd &= \sum_{i=1}^{n}\log  \int p(y_i|\bm{\psi})dp_{post}(\bm{\psi})=\sum_{i=1}^{n}\Big(\log[ E_{post} p(y_i|\bm{\psi} ) ] \Big)\\
				p_{waic} &= 2\sum_{i=1}^{n}\Bigg\{\Big(\log[ E_{post} p(y_i|\bm{\psi} ) ] \Big)-\Big(E_{post} \log[ p(y_i|\bm{\psi} ) ] \Big)\Bigg\}\\ \\
				WAIC &= 2\sum_{i=1}^{n}\Big(E_{post} \log[ p(y_i|\bm{\psi} ) ] \Big)-\sum_{i=1}^{n}\Big(\log[ E_{post} p(y_i|\bm{\psi} ) ] \Big)
			\end{split}\label{eq: WAIC_calc}
		\end{align}
		
		We have that $-2$ times the $WAIC$ in (\ref{eq: WAIC_calc}) brings it on the deviance scale. This allows for a more straight forward comparison with the AIC and DIC. We do not however do this in our case. For us, a larger WAIC value implies a better model fit.\\

		Here $\bm{\psi}$ is the unknown set of parameters. In our case we have that the unknown parameters is the set observation variances and the state variance. The integrations in (\ref{eq: WAIC_calc}) may be intractable since we have to integrate over the posterior distribution. However, the consistency of the sample means and the continuity of the $log$-function gives us that,
		\begin{align*}
			\begin{split}
				\log\Bigg(\frac{1}{S}\sum_{s=1}^S p(y_i|\bm{\theta}_s )\Bigg) \overset{p}{\rightarrow} \log\Big(E_{post}p(y_i|\bm{\theta})\Big)\\
				\Bigg(\frac{1}{S}\sum_{s=1}^S \log(p \big( y_i|\bm{\theta}_s )\big)\Bigg) \overset{p}{\rightarrow} E_{post}\log\Big(p(y_i|\bm{\theta})\Big)\\
			\end{split}
		\end{align*}
		
		This gives the computed \textit{WAIC} as,
		\begin{align*}
			\begin{split}
				cWAIC = 2\sum_{i=1}^{n}\Bigg(\frac{1}{S}\sum_{s=1}^S \log(p \big( y_i|\bm{\theta}_s )\big)\Bigg)-\\
				\sum_{i=1}^{n}\log\Bigg(\frac{1}{S}\sum_{s=1}^S p(y_i|\bm{\theta}_s )\Bigg)
			\end{split}
		\end{align*}
		
		and $cWAIC$ converges in probability to $WAIC$ as $S$ goes to infinity., where $S$ is the size of the posterior sample.\\
		
		\newpage
		\subsection*{Algorithms for Sampling}
		\subsubsection*{Algorithm 1: FFBS }

		\begin{algorithmic}[1]
			\State Given $\bm{m}_0,\bm{C}_0$, run the Kalman Filter.
			\State Draw $\te_n\sim N(\bm{m}_n,\bm{C}_n)$, where $\bm{m}_n,\bm{C}_n$.
			\For{$t$ in $t=n-1,...,0$}
			\State Draw $\te_t\sim N(\bm{h}_t,\bm{H}_t)$, ($\bm{h}_t,\bm{H}_t$ are defined in the main paper)
			\EndFor.

		\end{algorithmic}

		\subsection*{Algorithm 2: Posterior Sampling with FFBS step}
		\begin{algorithmic}[1]
			\State Initialize the unknown parameters, $\psi_1^{(0)},\psi_2^{(0)},...,\psi_p^{(0)}$
			\State Initialize state parameters using FFBS using, $\te^{(0)}_{0:n}\sim p(\te^{(0)}_{0:n}|\bm{y}_{1:n},\psi_1^{(0)},...,\psi_p^{(0)})$.
			
			\For{i in $1:nsims$}
			\State Draw $\psi_1^{(i)}\sim p(\psi_1^{(i)}|\bm{y}_{1:n},\te^{(i-1)}_{0:n},\psi_2^{(i-1)},...,\psi_p^{(i-1)})$
			\State Draw $\psi_2^{(i)}\sim p(\psi_2^{(i)}|\bm{y}_{1:n},\te^{(i-1)}_{0:n},\psi_1^{(i)},\psi_3^{(i-1)},...,\psi_p^{(i-1)})$
			\State $\vdots \quad\quad\quad\quad \vdots \quad\quad\quad\quad \vdots$
			\State  Draw $\psi_j^{(i)}\sim p(\psi_j^{(i)}|\bm{y}_{1:n},\te^{(i-1)}_{0:n},\psi_1^{(i)},...,\psi_{j-1}^{(i)},\psi_{j+1}^{(i-1)},...,\psi_p^{(i-1)})$
			\State Draw the state parameters using FFBS using, $\te^{(i)}_{0:n}\sim p(\te^{(i)}_{0:n}|\bm{y}_{1:n},\psi_1^{(i)},...,\psi_p^{(i)})$.
			\EndFor.
			
		\end{algorithmic}

		\clearpage

	\end{document}